\documentclass{lmcs}
\pdfoutput=1

\usepackage{lastpage}
\lmcsdoi{22}{2}{19}
\lmcsheading{}{\pageref{LastPage}}{}{}%
{Apr.~10,~2025}{May~20,~2026}{}

\usepackage{setspace}
\usepackage[utf8]{inputenc}
\usepackage[english]{babel}
\usepackage{amssymb}
\usepackage{placeins}
\usepackage{amsthm}
\usepackage{ marvosym }

\theoremstyle{definition}

\newtheorem*{problem*}{Decision Problem}
\usepackage{marvosym}
\usepackage{booktabs}

\usepackage{amsmath}
\usepackage{mathtools}
\usepackage{tabularx}
\usepackage{tikz-cd}
\usepackage{forest}
\usepackage{hyperref}
\hypersetup{
    colorlinks=true,
    linkcolor=blue,
    filecolor=magenta,      
    urlcolor=cyan
}

\usepackage{tikz}
\usetikzlibrary{automata, positioning, arrows, shapes}
\usetikzlibrary{shapes.geometric,calc,intersections}

\usepackage{mathtools}

\begin{document}

\title[Verification in Finite-Horizon Probabilistic Systems]{Verifying Equilibria in Finite-Horizon Probabilistic Concurrent Game Systems}

\author[S. Rajasekaran]{Senthil Rajasekaran\lmcsorcid{0000-0002-6675-8063}}
\author[M. Y. Vardi]{Moshe Y. Vardi\lmcsorcid{0000-0002-0661-5773}}

\address{Rice University Department of Computer Science}	
\email{sr79@rice.edu,vardi@rice.edu}  

 \begin{abstract}
 Finite-horizon probabilistic multi-agent concurrent game systems, also known as finite multiplayer stochastic games, are a well-studied model in computer science due to their ability to represent a wide range of real-world scenarios involving strategic interactions among agents over a finite number of iterations (given by the finite horizon). The analysis of these games typically focuses on evaluating (verifying) and computing (synthesizing/realizing) which strategy profiles (functions that represent the behavior of each agent) qualify as equilibria. The two most prominent equilibrium concepts are the \emph{Nash equilibrium} and the \emph{subgame-perfect equilibrium}, with the latter considered a conceptual refinement of the former. Computing these equilibria from scratch is, however, often computationally infeasible. Therefore, recent attention has shifted to the verification problem, where a given strategy profile must be evaluated to determine whether it satisfies equilibrium conditions. In this paper, we demonstrate that the verification problem for subgame-perfect equilibria lies in PSPACE, while for Nash equilibria, it is EXPTIME-complete.  This is a highly counterintuitive result since subgame-perfect equilibria are often seen as a strict strengthening of Nash equilibria and are intuitively seen as more complicated.
 \end{abstract}

\maketitle

\section{Introduction}

The ubiquity of distributed systems in modern computing architectures has motivated much recent work into the study of \emph{multi-agent systems}~\cite{GTW02,guptathesis,mogavero2014reasoning,SLmultiagentbook,van2008multi,Wool11,probsl,RBV23,RV22,RV21,bouyer2010nash,bouyer2011nash,Bouyer2014,BBMU15,raskinsubgame1,raskinsubgame2}, which are abstract mathematical models in which two or more rational agents interact with each other in order to achieve their goals. The broad motivation behind these works is to provide a mathematical framework for analyzing the overall trends of the complex mechanics that arise when multiple agents interact with one another in a strategic manner.

Much of the relevant work in this area comes from the field of \emph{formal methods}, which has recently begun to analyze equilibria in multi-agent systems as a way to analyze ``stable" behaviors in multi-agent systems, and, therefore, analyze the system's properties as a whole. The two major problems in the field of formal methods are the \emph{realizability}~\cite{PnuRos89a} and the \emph{verification} problems~\cite{Ros92,RV22}. Therefore, much of the work in formal analysis of multi-agent systems is concerned with determining the existence of (the realizability problem) and verifying (the verification problem) different types of equilibria~\cite{GTW02,bouyer2011nash,bouyer2010nash,BBMU15,RBV23,RV22,RV21,guptathesis,EVE,GPW17,BBMU15,raskinsubgame1,raskinsubgame2}.  As evidenced by the previously cited works, however, work in formal verification often comes with a strong preference for deterministic systems and deterministic agent behaviors. This creates something of an ideological divide between equilibrium analysis in the formal methods literature and the study of equilibria in the game theory literature.  Our broad goal in this paper is to begin to bridge this divide by applying this formal methods style of equilibrium analysis to probabilistic systems. This is a new line of inquiry, as even papers in the formal methods literature that consider probabilistic systems such as~\cite{GGLNWprob21} (which itself notes the focus on deterministic systems in previous literature) consider notions of almost-sure satisfaction on infinite-length executions of a system. These notions allow for these probabilistic systems to be analyzed through the same techniques used for deterministic ones.

In order to differentiate our work from this type of analysis, we consider finite-horizon systems, meaning that the system has an explicitly given finite bound on the number of steps in the execution of the system. This terminology follows the notation from the literature on Markov Decision Processes~\cite{Puterman94,etessamihandbook} and breaks from the terminology of the formal verification literature, where finite-horizon is usually shorthand for ``finite but unbounded"~\cite{Wool11,GV13}.  Finite-horizon systems are also arguably more popular than their infinite-horizon counterparts in the literature on Markov Decision Process~\cite{Puterman94,etessamihandbook,solan22,valit}, which can be seen as the ``single-agent" analog of the systems we consider in this paper.

We wish to consider the computational complexity of the realizability and verification problems in finite-horizon probabilistic multi-agent systems. The two most popular equilibrium concepts in the field of game theory are the \emph{Nash equilibrium}~\cite{Nash48} and the \emph{subgame-perfect equilibrium}~\cite{Osborne1994}, with the latter a strengthening of the former. However, this immediately presents a problem for the realizability problem. Since the realizability problem is concerned with determining whether an equilibrium exists, and we are considering probabilistic systems with a finite horizon, the realizability problem becomes trivial. This is because subgame-perfect equilibria can be shown to always exist in these systems through the use of algorithms like backwards induction~\cite{inductionnotes}. Furthermore, if we note that every subgame-perfect equilibrium is, by definition, a Nash equilibrium, then we note that Nash equilibria always exist in these systems as well.  If we remove the finite-horizon specification, then the status of the realizability problem for infinite-horizon systems now becomes a long-standing open problem in the field of game theory. In~\cite{chatterjee-majumdar-jurdzinski-csl2004}, it was claimed that there existed a 2-agent infinite-horizon game in which each of the agents had a reachability goal that did not have a Nash equilibrium. This claim, however, was shown to be flawed in~\cite{Bouyer2014}, so the question of whether the realizability problem is trivial in this setting is still open. Since even this is not known, the complexity of the realizability problem for Nash equilibria remains open in the infinite-horizon setting. Similarly, it is not known if a subgame-perfect equilibrium strategy profile always exists in the same unbounded setting. 

This leads us to focus on the verification problem for both the Nash and subgame-perfect equilibrium concepts. The verification problem takes an input of both a system and a candidate tuple of strategies and determines whether the input tuple of strategies satisfies the equilibrium condition.  In this way, it can naturally be seen as dual to the often considered problem of computing an equilibrium from scratch that is foundational to fields like algorithmic game theory~\cite{roughgarden2010algorithmic}.  However, despite the fact that recent works~\cite{RV22,ummelsthesis,parkerverification,KwiatkowskaNPS21} and tools~\cite{PRISM,EVE,CamaraGS015} have begun to consider practical aspects of the verification problem, there has been virtually no theoretical development of the verification problem in probabilistic settings.  

In order to accurately isolate the complexity of verification, we introduce a new succinctly represented model in this paper, the \emph{b-bounded} concurrent game system. In a b-bounded system, only b agents select an action concurrently at all given game-states, where b is a fixed constant. This model serves two purposes. First, it allows us to generalize from the popular but restrictive turn-based game model that is ubiquitous in the literature~\cite{ummelsthesis,guptathesis,Wool11,roughgarden2010algorithmic}. Second, it sharpens our complexity-theoretic results by circumventing the construction of an exponentially large transition table as part of our input. In a multi-agent system, the transition table of an unrestricted model can be seen as an exponential construction in itself, a point we discuss explicitly in Section~\ref{prelim} and that has been noted in other works such as~\cite{BBMU15}. This exponential construction within the input devalues the complexity-theoretic value of upper bounds, as a PSPACE upper bound would practically translate into doubly-exponential time complexity in the number of agents. We discuss how our results extend to a full concurrency model in Section~\ref{alternative} (one in which all agents select an action at all game-states).  Furthermore, the issue of goal representation also plays a significant role in the overall computational complexity of the verification problem. As noted in~\cite{RV21,RV22}, if agents are given highly succinct temporal logic formulae that represent their preferences on the system's executions, then the worst-case complexity of reasoning about these logics can completely overpower the worst-case complexity of the actual verification decision problem. In order to avoid this problem, we give the agents reachability goals, which are arguably the simplest non-trivial goals.

Thus, our stage is set. We are considering the computational complexity of the verification problem for both the Nash and subgame-perfect equilibrium in finite-horizon b-bounded game systems in which each agent has a reachability goal. Intuitively, if we recall that the subgame-perfect equilibrium is a conceptual strengthening of the Nash equilibrium, then we can expect that the associated verification decision problem for the subgame-perfect equilibrium is more difficult than the verification decision problem for the Nash equilibrium.

Our analysis starts with the subgame-perfect equilibrium. For the subgame-perfect equilibrium verification problem, we are able to show a PSPACE upper bound by employing results from circuit complexity and carefully considering issues of numerical representation that are often overlooked in the literature. As we demonstrate later on, a key subroutine in the verification decision problem is solving an exponentially-large system of equations. In~\cite{Puterman94}(Theorem 3.2, Corollary 3.2b), it is shown that systems of equations with rational coefficients admit rational solutions with polynomial-size encodings, meaning that for an exponentially-large system of equations, we would get an exponentially-large representation of a rational number solution in the worst-case. In this paper, we consider a numerical model directly inspired by the finite-precision arithmetic models used in real programs and explicitly construct a bound on solution sizes that is unique to our setting and model. These bounds allow us to get a PSPACE upper bound for the subgame-perfect equilibrium verification problem. If we note that backwards induction would compute a subgame-perfect equilibrium from scratch in EXPTIME, then this result lines up with our intuition, as pending a PSPACE=EXPTIME result, there is a computational benefit to verifying a given set of input strategies as opposed to explicitly synthesizing them.

We then move on to the Nash equilibrium verification problem, where our intuition is undone. Our results show that the problem of verifying a Nash equilibrium is EXPTIME-complete. Not only does this imply that checking if a strategy profile is a Nash equilibrium is asymptotically just as hard as computing a subgame-perfect strategy profile from scratch through backwards induction, but it also implies that the verification decision problem is easier for the stronger subgame-perfect equilibrium solution concept. At a high level, this can be understood by noting that the algorithm we present for the subgame-perfect equilibrium verification problem can be effectively parallelized, while the Nash equilibrium verification problem contains steps that are inherently sequential.

These results present an interesting relationship between these two ubiquitous equilibrium concepts and challenge our intuitive understanding of equilibria.

\section{Preliminaries}\label{prelim}
We start by introducing the main object of our study, \emph{b-bounded concurrent game system}. This model is very similar to the well-studied ``multi-player concurrent stochastic game'' model~\cite{solan22}. The major difference between our model and the most standard concurrent game models in the literature is the ``b-bounded'' property.

This ``b-bounded'' property is a limit on the concurrency of the game system, which is meant to strengthen our complexity theoretic results, a point we make explicit after introducing the formal definition.  By using the b-bounded model, we are able to provide an upper bound that is not inflated by the size of the transition function and a lower bound for the simplest type of b-bounded games -- 1-bounded games -- which are also known as turn-based games. 

\begin{defi}\label{pcgs}
A \emph{b-bounded concurrent game system} is a tuple $\mathbb{G} = \langle V, v_0, \Omega, A = \{A_1,A_2, \ldots A_{k}\},P, \delta,\mathbb{L} , G = \{G_1, G_2, \ldots G_{k}\} , F \rangle$ with the following interpretations

\begin{enumerate}
    \item $V$ is a finite set of states, with $v_0 \in V$ the initial state.
    \item $\Omega = \{1,2, \ldots k\}$ is the set of agents. We denote the size of $\Omega$ by $k$, i.e., the system has $k$ agents.
    \item $A = \{A_1, A_2, \ldots A_k\}$ is the set of finite action sets, where Agent $i$ is associated with the set of actions $A_i$. We refer to cross-product spaces $\Theta_W = \bigtimes_{i \in W} A_i$ for some $W \subseteq \Omega$ and then use $\theta$ to refer to an element of $\Theta_W$.
    \item $P : V \rightarrow 2^{\Omega}$ is the \emph{playing function}. This function decides which agents are ``active" at a given state, as not every agent makes a relevant decision at every state. The game being $b$-bounded means that at most $b$ agents can make a relevant decision at all states, i.e., $|P(v)| \leq b$ for all $v \in V$ for $b\geq 1$.
    \item $\delta$ is the transition function. Let $v, v'$ be elements of $V$ and let $\theta$ be an element of $\Theta_{P(v)}$. The transition function $\delta$ maps triples $(v,\theta,v')$ to  $[0,1]$. These outputs represent a probability distribution, and we therefore have $\sum_{v' \in V} \delta(v, \theta, v') =1$. Intuitively, the transition function only considers the actions of agents that are included in the image of the playing function when considering possible transitions from a state. Note that it is possible that $P(v) = \emptyset$. In this case, we say that $v$ is \emph{uncontrolled} as the transitions from state $v$ are determined by a fixed probability distribution that does not take input from the agents, analogous to uncontrollable ``probabilistic" states in the theory of Markov Decision Processes~\cite{Puterman94}.
    \item In order to deal with issues of numerical representation, we assume that all probabilities output by $\delta$ are represented by decimals that are exactly $\mathbb{L}$ bits long, where $\mathbb{L}$ is provided in unary.     
    \item $G = \{ G_1, G_2, \ldots G_k\}$ is a set of \emph{reachability goals}. The goal $G_i \subseteq V$ represents the reachability goal of Agent $i$.
    \item $F \in \mathbb{N}$ is a finite time horizon written in binary.
\end{enumerate}
\end{defi}

Note that in this definition, we have $1$-indexed the set of agents as opposed to $0$-indexing that is more present in the related computer science literature. This is because the model presented more closely lines up with models studied in game theory~\cite{solan22,Fudenberg13}, and the game theory literature adopts mathematical conventions such as $1$-indexing as opposed to computer science conventions such as $0$-indexing. 

Taken together, our concurrent game model represents a very well-studied standard finite-horizon stochastic game model~\cite{solan22} with one exception: the aforementioned b-bounded property. In this model, we have a bound on the concurrency of the system given by the playing function $P$ and the bound $b$. Note that in order for this model to be considered a ``game" where agents make decisions, we must have $b \geq 1$. This bound was introduced for the purpose of complexity-theoretic analysis, as, without this bound, the transition table $\delta$ can be seen as an exponential construction itself. For example, if we have $k$ agents and each agent has two actions, a transition table would need to have $|V|^2 \times 2^k$ entries in the worst-case scenario. This table would represent the probability that a state $v_1$ transitions to a second state $v_2$ under each of the $2^k$ collective action choices in the full action product space $\Theta_\Omega$.  The construction of such a large transition table would devalue the meaning of the complexity-theoretic results presented in this paper, as an exponential time result in the size of the game input would be interpreted as doubly-exponential in the number of agents. This led us to choose a natural restriction of concurrent games that would have a transition table with a more succinct presentation. In a b-bounded game, the transition table has at most $|V|^2 \times 2^b $ entries for a fixed $b$ (and therefore polynomial in the number of agents). This allows us to demonstrate upper bounds that hold even for models that have a more compact transition table representation. Furthermore,  the new size bound allows us to better tie our complexity results to the number of agents and states in the game, as the transition table no longer adds an extra exponential blowup to our complexity-theoretic analysis. In Section~\ref{alternative}, we explicitly consider the upper bound for a full concurrency model and show that the PSPACE membership result of Theorem~\ref{upper} still holds for this model as well.

The game starts at the state $v_0$. At this state, the agents in $P(v_0)$ choose a distribution over the actions from their action set. This collective choice is then input into the transition function $\delta$, which returns a distribution over a set of new states. This distribution is then sampled, and the execution of the system proceeds to the next state $v'$, where the process repeats with the agents in $P(v')$ until $F$ states have been visited (it is acceptable to visit a single state multiple times). After the $F-1$-th action choice, the game terminates, and no further moves are made. The entire record of the $F-1$ action choices and the corresponding states visited is called a \emph{play}.

Agents are incentivized to choose their actions so that a state in their reachability goal is eventually seen. In other words, agents prefer plays that visit a state in their reachability goal to those that do not.  For an Agent $i$, a play that visits a state in its reachability set yields a \emph{payoff} of $1$, whereas a play that does not visit a state in its reachability set yields a payoff of $0$. Agents select their actions to maximize their expected payoffs.

Agents choose their actions based on the observed history of the game, which we model as an element of $V^+$. Therefore, a \textbf{strategy} for Agent $i$ is a function of type $V^+\rightarrow \mathbb{D}(A_i)$, where the notation $\mathbb{D}$ is used to refer to a distribution over a set.  We use the notation $\pi_i$ to refer to a strategy for Agent $i$ and the notation $\Pi_i$ to refer to the set of all strategies for Agent $i$, i.e. the set of functions of type $V^+\rightarrow \mathbb{D}(A_i)$.  Strategies are mathematically modeled through the use of \emph{probabilistic transducers}, the definition of which is given with respect to a model for an Agent $i$ strategy $\pi_i$.

\begin{defi}\label{probabilistictransducermodel}
A \emph{probabilistic transducer} is a 5-tuple $\langle S_i , s^i_0 , V, \gamma^i, O_i \rangle$, where
\begin{enumerate}
\item $S_i$ is a finite set of states, and $s^i_0 \in S_i$ is the initial state.
\item $V$, the set of states in the corresponding concurrent game model, is the input alphabet.
\item $\gamma^i : S_i \times V \rightarrow S_i$ is the deterministic transition function.
\item $O_i : S_i \times V \rightarrow \mathbb{D}(A_i)$ is the probabilistic output function.
\end{enumerate}
\end{defi}

In our model, probabilistic transducers have a deterministic transition function and a finite state space.  A discussion of why considering such transducers is sufficient and what would happen if a probabilistic transition function were used instead can be found in Section~\ref{alternative}. Since transducers are provided as part of our input, we once again assume that all probabilities output by $O_i$ are exactly $\mathbb{L}$  (by slight abuse of notation, $\mathbb{L}$ was defined in $\mathbb{G}$) bits in length to deal with issues of numerical representation. Unlike the deterministic transition function and the finite state space, this is a restriction on the model as it does not allow us to represent repeating decimals or irrational numbers. For repeating decimals, we note that the numbers do not necessarily have to be represented in base two, though we assume this for our computations. For irrational numbers, there exist games in which the only equilibria involve distributions with irrational probabilities~\cite{Nash51}. Nevertheless, the verification problem is ultimately a problem about programs, and real programs can only work with finite precision, so a finite representation of probabilities is a reasonable restriction.  A further extension of this work would be to consider rational-number representations of probabilities, which can be seen as another ``finite" representation. This would introduce new issues with numerical representation about how to represent and perform operations on the rational representations of the probabilities.

The input alphabet of a probabilistic transducer is $V$, meaning that the probabilistic transducer can run on elements of $V^+$. 
The probabilistic transducer thus represents functions of the type $V^+ \rightarrow \mathbb{D}(A_i)$, i.e., strategies. The output function has domain type $S_i \times V$, so when Agent $i$ is asked to output an action distribution at a state $v$ such that $i \not \in P(v)$, we can specify that the output is a special character $\bot$. This allows us to account for the restrictions of the playing function $P$ in $\mathbb{G}$ with minimal notational overhead.

\begin{defi}\label{product}
A \emph{strategy profile} $\pi$ is a tuple of individual agent strategies $\langle \pi_1 \ldots \pi_{k} \rangle $ represented by probabilistic transducers. It is represented by the  \emph{product transducer} $T_{\pi} =\langle S, s_0, V, \gamma, O \rangle$ is a probabilistic transducer defined in the following way:
\begin{enumerate}
 \item  The set of states $S$ is given by a cross product of the component state-sets, $ S = \bigtimes_{i \in \Omega} S_i$.   $s_0 \in S$ is given by $\langle s^0_0, \ldots, s^{k-1}_0 \rangle$.
\item $V$, the input alphabet remains unchanged.
\item The transition function $\gamma: S \times V \rightarrow S$ works component-wise. The state $\langle s^1, \ldots, s^{k} \rangle \in S$ transitions to the state $\langle \gamma_0 (s^1, v) \ldots \gamma_{k-1}(s^{k}, v) \rangle$ upon reading  $v \in V$.
\item  The output function $O$ aggregates the component-wise output functions. The output $O(\langle s^1, \ldots ,s^{k} \rangle, v)$ is given by $\langle O_1(s^1,v),$ $ \ldots, O_{k}(s^{k},v) \rangle$. For $i \in P(v)$, Each component $O_i(s^i,v)$ belongs to $\mathbb{D}(A_i)$. For $j \not \in P(v)$, the output $O_i(s^j,v)$ is $\bot$. By a slight abuse of notation, we project these $\bot$ values out, and so, taken together, $O(s,v) \in \bigtimes_{i \in P(v)} \mathbb{D}(A_i)$.
\end{enumerate}

This construction $T_{\pi}$ is useful for representing a tuple of agent strategies as a single transducer. Although it may not be the case that all agent actions need to be considered at every state due to the playing function $P$, this formulation contains enough information to deal with all possible playing functions.

\end{defi}

A strategy profile induces a probability distribution over the set of possible plays (since a single play is finite due to $F$). Given a strategy profile, we can associate a payoff to each agent by computing the expected value of the plays generated over the distribution. 

\begin{defi} The payoff for Agent $i$ under a strategy profile $\pi$ denoted $p(\pi,i)$ can be calculated    
$ p(\pi,i) =  \sum_{t \in \mathrm{Plays}} \mathbb{P}(\pi,t) * \mathrm{Payoff}(t,i)$ where  $\mathbb{P}(\pi,t)$ refers to the probability that when all agents follow $\pi$ the play $t$ is sampled and  $\mathrm{Payoff}(t,i)$ is the payoff Agent $i$ receives from the play $t$ ; 1 if it contains  a state in $G_i$ and $0$ otherwise.
\end{defi}
Given this notion of a payoff under a strategy profile, we can now define our first solution concept of interest.

\begin{defi}    
Let $\pi = \langle \pi_1 \ldots \pi_i \ldots \pi_{k} \rangle$
be a strategy profile.  $\pi$ is a {\em Nash equilibrium} iff for every Agent $i$ and for all strategy profiles of the form   $\pi' = \langle \pi_1 ,\ldots, \pi_{i-1}, \pi'_i, \pi_{i+1}, \ldots, \pi_{k} \rangle$, where $\pi'_i$ is an arbitrary element of $\Pi_i$, 
we have $p(\pi, i) \geq p(\pi' , i)$ 
(otherwise, we say Agent $i$ has a \emph{profitable} deviation).\end{defi} 

In other words, a Nash equilibrium is a strategy profile such that any unilateral change in strategy for any Agent $i$, from $\pi_i$ to some other $\pi'_i$,  does not result in a greater probability that Agent $i$'s goal is satisfied. Therefore, Agent $i$ is not incentivized to change her strategy unilaterally. The \textbf{Nash-equilibrium verification problem} takes a concurrent game $\mathbb{G}$ and a strategy profile $\pi$ specified by a tuple of probabilistic transducers  $\langle \pi_1 \ldots \pi_i \ldots \pi_{k} \rangle$ as inputs, and the problem is to decide whether $\pi$ is a Nash equilibrium in $\mathbb{G}$.

The Nash equilibrium is not the only solution concept studied in this paper. In order to introduce the second, we first introduce the notion of a \emph{history}.

\begin{defi}
    A \emph{history} $h \in V^+$ is a finite set of states satisfying the following three requirements.
    \begin{enumerate}
        \item The first element of $h$ is $v_0$, the initial state.
        \item For each pair of consecutive elements $h[j], h[j+1]$ in $h$, it must be the case that there exists some tuple of actions that transitions from $h[j]$ to $h[j+1]$ with positive probability, i.e., $ \exists \theta \in \bigtimes_{i \in P(h[j])} A_i. \delta(h[j],\theta,h[j+1]) >0$.
        \item $|h| \leq F$.
    \end{enumerate}
\end{defi}

A history intuitively represents a partial execution of a play (and so cannot be longer than a play itself).

\begin{defi}
    Given a strategy $\pi_i$ for Agent $i$ and a history $h$, the \emph{substrategy} $\pi_i|_h$ is defined as the function 
    $\pi_i|_h(w) := \pi_i(h \cdot w)$ for all $w \in V^*$.
\end{defi}

Intuitively, the sub-strategy acts like the strategy $\pi$ that has already witnessed the history $h$. Since our strategies are represented by finite-state transducers with deterministic transition functions, creating a substrategy is as simple as updating the initial state of the transducer to the state that results after reading $h$.

\begin{defi}
    Given a concurrent game $\mathbb{G}$ and a history $h$. the subgame $\mathbb{G}|_h$ is the game $\mathbb{G}$ after having observed the history $h$. Its initial state is the last element of $h$, and depending on the contents of $h$, some agents might have a fixed payoff of $1$ in this subgame if $h$ visited one of their goal states. Furthermore, the finite horizon of $\mathbb{G}|_h$ is updated to $F - |h|$, where $F$ is the original finite horizon of $\mathbb{G}$.  
\end{defi}

A subgame is just the game after having witnessed the history $h$. Therefore, the initial state changes along with the computation of the payoff function; agents that have already achieved their goal in the history receive a fixed payoff of $1$.

\begin{defi}\label{sgpedef}
    A strategy profile $\pi = \langle \pi_1, \pi_2, \ldots \pi_k \rangle$ is a \emph{subgame-perfect equilibrium} in a concurrent game $\mathbb{G}$ iff it is the case that for all histories $h$ we have that $\pi|_h := \langle \pi_1|_h, \pi_2|_h, \ldots \pi_k|_h \rangle$ is  a Nash equilibrium in $\mathbb{G}|_h$.
\end{defi}

A \emph{subgame-perfect equilibrium} is a strategy profile $\pi$ that satisfies the Nash equilibrium condition after every history. This means that for every plausible history $h$ we have that $\pi|_h$ is a Nash equilibrium in $\mathbb{G}_h$. Note that histories are not dependent on $\pi$, as histories correspond to paths of states that occur with positive probability when arbitrary agent behaviors are considered, not just the behavior suggested by $\pi$.  This is opposed to Nash equilibrium, which can be thought of as verifying the equilibrium condition for the starting state history $h = v_0$ only. The subgame-perfect equilibrium is then naturally thought of as strengthening the Nash equilibrium. The \textbf{subgame-perfect-equilibrium verification problem} takes as input a concurrent game $\mathbb{G}$ and a strategy profile $\pi$ and decides whether $\pi$ is a subgame-perfect equilibrium in $\mathbb{G}$.

\section{subgame-perfect Equilibrium Upper Bound}\label{subgameperfect}

In this section, we prove that the subgame-perfect equilibrium verification problem can be decided in PSPACE. In order to do so, we construct a Markov Chain that keeps track of the relevant information needed to analyze the behavior of the strategy profile $\pi$ in $\mathbb{G}$. By keeping track of the current state of $\mathbb{G}$, the state space of all strategies $\pi_i$ in the tuple $\pi$, and the current time index, we capture all of the relevant information needed to analyze the behavior of $\pi$ in $\mathbb{G}$.

Given a concurrent game $\mathbb{G}$ and an input strategy profile $\pi$, we construct the Markov Chain $\mathbb{G} \times \pi$, which is the cross product of $\mathbb{G}$ and $\pi$. Formally,
$\mathbb{G} \times \pi = \langle V \times S \times [F], P \rangle$ in the following way. The state space is given by the cross product of the sets $V$ (the state set of $\mathbb{G}$), $S$ (the state set of the cross product transducer $T_{\pi} =\langle S, s_0, V, \gamma, O \rangle$ given in~ Definition~\ref{product}) and $[F]$ (the set $\{1,2, \ldots F \}$, where $F$ is the finite time horizon of $\mathbb{G}$).  Since $S$ is the state space of $T_{\pi}$, the cross product of all strategies in $\pi$, and $F$ was originally provided in binary in $\mathbb{G}$; this state space has exponential size w.r.t. the input of $\mathbb{G}$ and $\pi$. Elements of the transition matrix $P$ are defined as follows. Given two elements $\langle v_1,s_1,n_1 \rangle, \langle v_2,s_2,n_2 \rangle \in V \times S \times [F]$, the probability associated from transitioning from $\langle v_1,s_1,n_1 \rangle$ to $\langle v_2,s_2,n_2 \rangle$ is the product $p_v \cdot p_s \cdot p_n$ of three different components: $p_v$, the probability that $v_1$ transitions to $v_2$, $p_s$, the probability that $s_1$ transitions to $s_2$, and $p_n$, the probability that $n_1$ transitions to $n_2$.

We first discuss $p_s$. Recall that $s_1$ and $s_2$ are states of the product transducer $T_\pi$. Therefore, we expect a state $s_1 = \langle s_1^1 \ldots s_1^k \rangle$ to transition, upon seeing $v_1$,  to $s = \langle \gamma^1( s_1^1, v_1) \ldots \gamma^k(s_1^k, v_1) \rangle$, where $\gamma^i$ is the transition function of $\pi_i$. If $s_2$ matches $s$ then  $p_s = 1$, otherwise  $p_s = 0$.

Next, we consider $p_n$. Quite simply, $n_2$ should be $n_1 + 1$, representing the time-step incrementing. Generally, $p_n = 1$ if $n_2 = n_1 + 1$ and $p_n =0$ otherwise. If, however, $n_1 = F$, we say that the state $\langle v, s, F \rangle$ has no outgoing transitions, so the state is a terminal state. This is representative of the game terminating after $F$ time steps.

Finally, we consider $p_v$. Intuitively, we want the probability that $\mathbb{G}$ transitions from $v_1$ to $v_2$ under the action distribution $O(s_1)$. Formally, we have
$$ \sum_{\theta \in supp(O(s_1,v_1))} \mathbb{P} ( O(s_1,v_1)[\theta]) \cdot \delta(v_1, \theta, v_2)$$
where $supp(\cdot)$ refers to the \emph{support} of a probabilistic distribution, i.e., the set of elements that are assigned a non-zero probability by the distribution. 
The expression $\mathbb{P} ( O(s_1,v_1)[\theta])$ represents the probability that, when sampled, $O(s_1,v_1)$ returns $\theta$. In our bounded-concurrency setting, this sum iterates over $\bigtimes_{ i \in P(v_1)} A_i$ -- details about the full-concurrency setting can be found in Section~\ref{alternative}.  This set has polynomial size w.r.t the input due to the bound $b$, so it can be iterated through in PSPACE.

Since $p_v$ can be computed in PSPACE, the product $p_v \cdot p_s \cdot p_n$ can be computed in PSPACE. This means that it is possible to query elements of the transition table $P$ in PSPACE.  

\begin{lem}\label{transprobs}
Given two states $s,r \in \mathbb{G} \times \pi$,  it is possible to compute the transition probability $P(s,r)$ in PSPACE w.r.t to the size of the input of $\mathbb{G}$ and $\pi$
\end{lem}

\subsection{Outline}\label{outline}

\begin{figure}[ht]
  \centering
  \includegraphics[width=\textwidth]{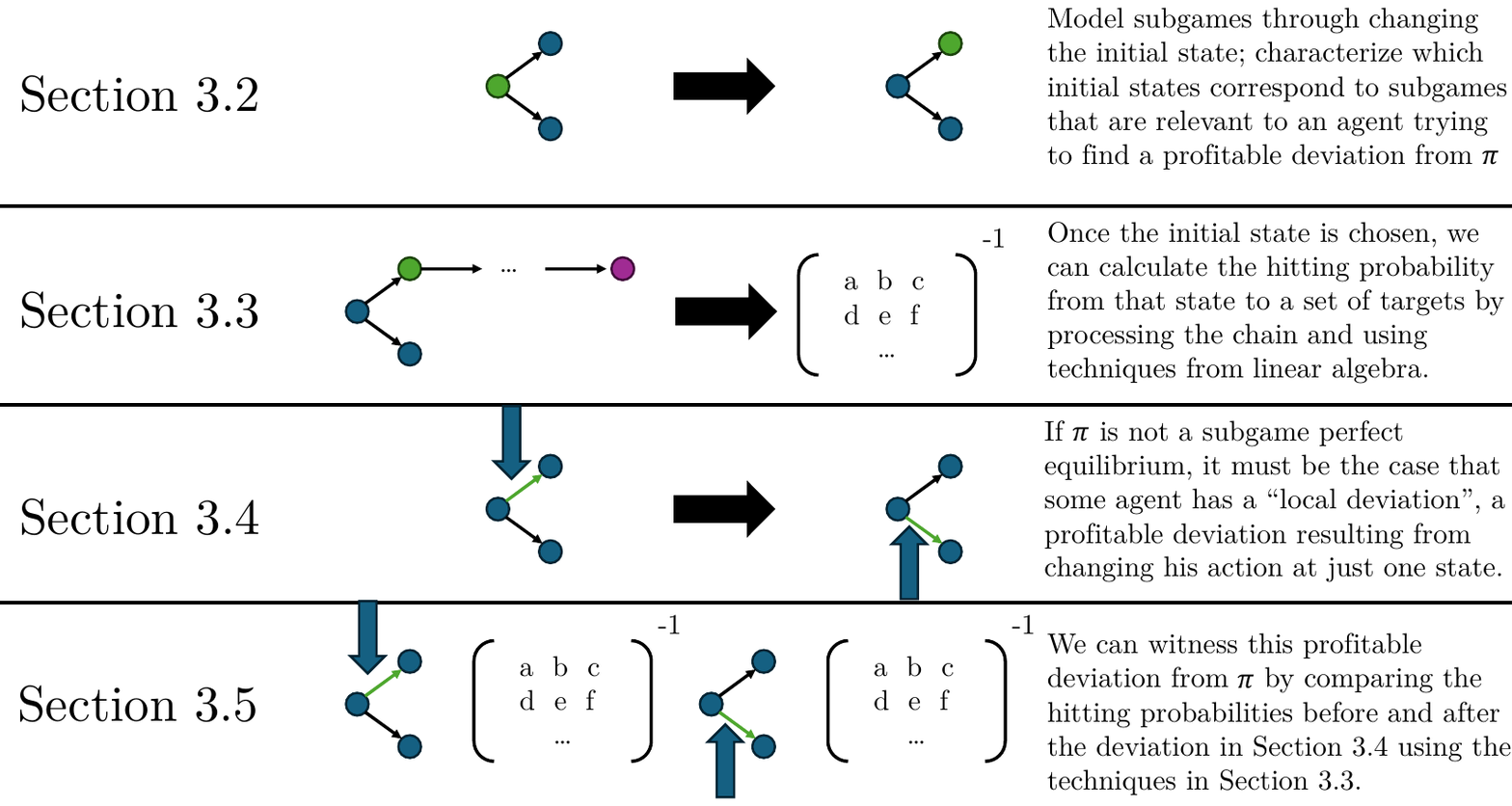}
  \caption{A chart depiction of the outline with some figures intended to facilitate conceptual understanding.}
  \label{fig:outlinesec3}
\end{figure}

The overarching goal of this section is to prove that the problem of deciding whether a strategy profile $\pi$ is a subgame-perfect equilibrium in a $b$-bounded concurrent game $\mathbb{G}$ is in PSPACE. This consists of several steps, which are detailed in the following sections, and a broad overview of these sections and their goals can be found in Figure~\ref{fig:outlinesec3}.  In order to establish our PSPACE upper bound, we first recall that PSPACE is closed under non-determinism and complementation, i.e., PSPACE=NPSPACE=co-PSPACE. Since PSPACE is closed under complementation, we consider the problem of determining whether $\pi$ is \emph{not} a subgame-perfect equilibrium in $\mathbb{G}$.

By Definition~\ref{sgpedef}, $\pi$ is a subgame-perfect equilibrium in $\mathbb{G}$ if, for all histories $h$, $\pi|_h$ is a Nash equilibrium in $\mathbb{G}|_h$. The negation of this condition then states that there exists some history $h$ such that $\pi|_h$ is not a Nash equilibrium in $\mathbb{G}|_h$, meaning that there exists some Agent $i \in \Omega$ that can achieve a higher payoff by deviating from $\pi|_h$.  For the rest of this section, we fix our analysis to the Agent $i$, noting that if it is possible to check whether a single Agent $i$ has a profitable deviation in PSPACE, then it is possible to guess the agent to check using non-determinism since NPSPACE=PSPACE.

The first step to our approach is to show that we can compute the probability that Agent $i$ meets his goal in $\mathbb{G}|_h$,
when all the agents follow $\pi|_h$ in NPSPACE, for all histories $h \in V^+$.  This represents the probability that the agent gets in $\mathbb{G}|_h$ if he does not deviate. In Section~\ref{historysubgame} we discuss how we can determine whether or not a state in $\mathbb{G} \times \pi$ can be reached by following a history, allowing us to analyze the set of subgames in $\mathbb{G}$ through the set of states reachable by some history in $\mathbb{G} \times \pi$. In Section~\ref{computation}, we show how to compute the probability that Agent $i$ reaches his goal from a state in $\mathbb{G} \times \pi$ in PSPACE. 

Then, we consider deviations. In Section~\ref{improve}, we prove that if $\pi$ is not a subgame-perfect equilibrium in $\mathbb{G}$, then a ``special" history must exist and can be guessed in NPSPACE. After this special history $h$ is witnessed, the agent only needs to make a ``small" change to $\pi_i|_h$ 
in order to immediately improve his probability of reaching his goal in $\mathbb{G}|_h$. Since this change is small, we can reapply the techniques employed in Section~\ref{computation} to recompute the probability that the agent now receives under this alternate strategy. This is the major difference between the subgame-perfect-equilibrium-verification problem and the Nash-equilibrium verification problem. The Nash-equilibrium verification problem only considers the starting state history $h = v_0$, and it is not guaranteed that this history has this special property.

After the agent changes his strategy in $\mathbb{G}|_h$ for the special history $h$, we are now given two probabilities -- the probability of reaching the goal before the change and the probability after. In order to prove that $\pi$ is not a subgame-perfect equilibrium, it must be the case that the latter is strictly larger than the former. The final task is then to compare these two probabilities in PSPACE, which is handled in Section~\ref{compare}.  Finally, questions about the full concurrency model and our model of strategies as transducers with deterministic transition functions are answered in Section~\ref{alternative}.

\subsection{Histories and Subgames}\label{historysubgame}
As outlined in Section~\ref{outline}, the first step to our approach is to guess a history $h$ so that we can check whether $\pi|_h$ is a Nash equilibrium in $\mathbb{G}|_h$. Since our analysis is based on the Markov Chain $\mathbb{G} \times \pi$, we are interested in the set of states that result from following an arbitrary history $h$.

\begin{defi}\label{historyreldef}
    A state $r \in V \times S \times [F]$ belongs to the set $R$ of \emph{relevant history-reachable states} iff there exists a history $h$ that reaches $r$ from $\langle v_0, s_0 ,1 \rangle$. This means that there is a path of states from $\langle v_0, s_0 , 1 \rangle$ to $r$  such that for each pair of consecutive states $\langle v_j, s_j , n_j \rangle,  \langle v_{j+1}, s_{j+1} , n_{j+1} \rangle$ we have $\gamma(s_j, v_{j}) = s_{j+1}$, $n_{j+1} = n_j + 1$  and there exists a global action  $\theta  \in \bigtimes_{k \in P(v_j)} A_k $ such that $\delta(v_j,\theta, v_{j+1}) >0$. Furthermore, for all elements $\langle v_j, s_j , n_j \rangle$ in the path we have $v_j \not \in G_i$. 
\end{defi}

The specification that no element in the path contains an element with $v \in G_i$ is made since our analysis fixes the Agent $i$. Since we are searching for histories where Agent $i$ has a profitable deviation, we do not want to include histories $h$ that already visit $G_i$ since the Agent $i$ would then receive a payoff of  $1$ in $\mathbb{G}|_h$ no matter what and would therefore not have a profitable deviation in $\mathbb{G}|_h$ by definition.

\begin{lem}\label{historyrellemma}
The problem of deciding whether a state $r = \langle v_r, s_r, n_r \rangle \in V \times S \times [F]$ belongs to $R$ is in PSPACE.
\end{lem}
\begin{proof}
    Intuitively, the idea is that this property can be determined through a reachability query that simulates the progression of an arbitrary history obtained from choosing arbitrary actions. Starting from the state $\langle v_0, s_0, 1 \rangle$, a path from $\langle v_0, s_0 , 1 \rangle$ to $\langle v_r, s_r, n_r \rangle$ can be guessed state by state on-the-fly in NPSPACE=PSPACE. At each state $\langle v_j, s_j, n_j \rangle$, the algorithm can guess a successor $\langle v_{j+1}, s_{j+1}, n_{j+1} \rangle$ and verify in PSPACE that $ \gamma(s_j, v_{j}) = s_{j+1}$, $n_{j+1} = n_j + 1$  and there exists a global action  $\theta  \in \bigtimes_{k \in P(v_j)} A_k $ such that $\delta(v_j,\theta, v_{j+1}) >0$,  by iterating over all possibilities for $\theta$. It can also verify that $v_j \not \in G_i$. The algorithm only needs to store at most two states in its memory -- the current state and the guessed successor. Since the state space $V \times S \times [F]$ is exponential in the size of the input, individual states can be encoded in PSPACE. 
\end{proof}

The set $R$ of relevant history-reachable states represents the states that occur in $\mathbb{G} \times \pi$ after some history $h$ in $\mathbb{G}$ has been observed. Therefore, if we want to analyze the payoff given to Agent $i$ in a subgame induced by a history $h$ in $\mathbb{G}$, we consider $\mathbb{G} \times \pi$ started from the state $r \in R$ that results from following the history $h$. 

\subsection{Computing Winning Probabilities From a State}\label{computation}
Following our outline, the next step is to compute the probability that Agent $i$ reaches his goal in $\mathbb{G}|_h$ when all agents (including $i$) follow $\pi|_h$ for some history $h$. Lemma~\ref{historyrellemma} gives us a way to guess an arbitrary history in PSPACE. Therefore, we assume that both the Agent $i$ and the history $h$ are fixed in this section and denote the relevant history-reachable state resulting from $h$ in $\mathbb{G} \times \pi$ as $r$.   Starting from the state $r \in R$  in $\mathbb{G} \times \pi$, we want to compute the probability that a member of the set $T = \{ \langle v,s,n \rangle | v \in G_i\}$ is reached, as this is equivalent to the probability that Agent $i$ reaches a state in his goal set in $\mathbb{G}|_h$ when the strategy $\pi|_h$ is followed. In the Markov Chain $\mathbb{G} \times \pi$, the probability that a member of some set is visited when starting from some state is called a \emph{hitting probability}~\cite{Puterman94}, and so we are interested in the hitting probability $h_r$ of $T$ when started in state $r$.

A standard way to solve for a hitting probability is to create a system of linear equations~\cite{Puterman94}. In $\mathbb{G} \times \pi$, for states $t \in T$, we would use the equation $h_t = 1$ -- a state already in $T$ visits $T$ with probability $1$.  For other states $r$, we would use the equation $h_r = \sum_{v \in V \times S \times [F]} P(r,v) \cdot h_v$ -- this relates the hitting probability of $r$ to the hitting probabilities of the states its transitions to. Taken together, the hitting probabilities would be computed by a set of $|V \times S \times [F]|$ equations over $|V \times S \times [F]|$ states.  However, it is possible for this system to be underdetermined, meaning that standard linear algebra techniques would not apply. To demonstrate this, we abstract away from our setting and illustrate how a standard linear-algebraic approach could go wrong in Example~\ref{markovexamplehittingprobs}. This example is intended to inform readers unfamiliar with the application of linear-algebraic techniques to Markov Chains of the general setting; we then discuss why this potential error does \emph{not} apply to our setting.  

\begin{exa}\label{markovexamplehittingprobs}
    Consider a Markov Chain $\mathcal{C}$ with two non-target states $x$ and $y$ with the following properties: The state $x$ self-transitions with probability $\frac{1}{2}$ and transitions to $y$ with probability $\frac{1}{2}$. Furthermore, the state $y$ self transitions with probability $\frac{1}{2}$ and transitions to $x$ with probability $\frac{1}{2}$ -- see Figure~\ref{fig:example} for an illustration. The equations for the hitting probabilities $h_x$ and $h_y$ would be both given by the linear equation $\frac{1}{2}h_x + \frac{1}{2}h_y$, meaning that the system of equations used to solve for hitting probabilities in $\mathcal{C}$ would be underdetermined and have infinite solutions. 
\end{exa}
\begin{figure}
    \centering
    \begin{tikzpicture}[>=stealth, node distance=3cm, every loop/.style={min distance=10mm}]

    \begin{scope}
        \filldraw[fill=gray!20, draw=gray!50, thick]
            (-4,2.5) .. controls (-3.5,3.5) and (-2.5,3.5) .. (-2,2.8)
            .. controls (-1.8,2.3) and (-2.5,1.5) .. (-3,1.6)
            .. controls (-3.7,1.5) and (-4.2,2.0) .. (-4,2.5)
            -- cycle;
        \node at (-3,2.5) {\Large \textbf{$\mathcal{C'}$}};
    \end{scope}

    \node[state] (x) {$x$};
    \node[state, right=of x] (y) {$y$};

    \path[->]
        (x) edge[loop above] node{$\tfrac{1}{2}$} (x)
        (y) edge[loop above] node{$\tfrac{1}{2}$} (y)
        (x) edge[bend left, above] node{$\tfrac{1}{2}$} (y)
        (y) edge[bend left, below] node{$\tfrac{1}{2}$} (x);

\end{tikzpicture}

    \caption{An illustration of Example~\ref{markovexamplehittingprobs}. The entire figure represents the Markov Chain $\mathcal{C}$, which has two disconnected components:  $\mathcal{C'}$ and the two non-target states $x$ and $y$. We wish to calculate the hitting probabilities for some state $c \in \mathcal{C'}$ by setting up a system of linear equations. If we set the hitting probabilities up in a na\"ive way, we get $h_x = h_y = \frac{1}{2}h_x + \frac{1}{2}h_y$, a pair of equations with an infinite number of solutions. Thus, the system of equations to solve for the hitting probabilities in $\mathcal{C}$ is underdetermined. }
    \label{fig:example}
\end{figure}

The Markov Chain $\mathbb{G} \times \pi$, however, has a special structure. Since it explicitly tracks the time index $n \in [F]$, this means that it contains no cycles -- states with time index $n$ can only transition to states that have time index $n+1$. 

\begin{lem}\label{uniquesolution}
    The system of equations used to solve for the hitting probabilities in $\mathbb{G} \times \pi$ has a unique solution.
\end{lem}
\begin{proof}
    We demonstrate this by backwards induction. Consider the set of states $\langle v,s,F \rangle$, i.e. the set of states with time index $n = F$. These states have no outgoing transitions, so we have $h_{\langle v,s,F \rangle} = 1$ if $v \in G_i$ and $0$ otherwise. Therefore, all states of the form $\langle v,s,F \rangle$ have unique solutions for their hitting probabilities.

    Assume now that all states of the form $\langle v,s,n+1 \rangle$ have unique solutions for their hitting probabilities. We then show how to uniquely determine the hitting probabilities of states of the form  $\langle v,s,n \rangle$. For states $\langle v,s,n \rangle$ with $v \in G_i$, we have $h_{\langle v,s,n \rangle} = 1$. Otherwise, we can express $h_{\langle v,s,n \rangle} = 1$ as a linear combination of some $h_{\langle v,s,n+1 \rangle}$, which are uniquely determined by the induction hypothesis. Therefore, $h_{\langle v,s,n \rangle}$ is uniquely determined as well.
\end{proof}

Solving a system of explicitly given linear equations with a unique solution belongs to the class NC\textsuperscript{2}~\cite{csanky76}, a subset of the class NC of problems solvable in parallel poly-logarithmic time using polynomially many processors~\cite{cook85}. The standard solution concept for a system of linear equations is a vector ${\bf v}$ that associates a numerical value to each variable $v$ in the system of equations. Since NC\textsuperscript{2} is a complexity class containing decision problems~\cite{johnson-chapter-handbook-of-tcs}, the algorithm in~\cite{csanky76} technically takes as input a variable $v$ and an index $k$ and decides the value of $k$-th bit of  $v$ in ${\bf v }$.
The class NC\textsuperscript{2} itself contained in  log\textsuperscript{2}-SPACE~\cite{johnson-chapter-handbook-of-tcs}, meaning that we can create a SPACE($\log^2 n)$ Turing Machine $\mathcal{T}$ that is capable of simulating the NC\textsuperscript{2} algorithm presented in~\cite{csanky76}. The machine $\mathcal{T}$ takes a variable $v$, and an index $k$ as input, and it simulates the relevant processors of the NC\textsuperscript{2} algorithm to decide whether the $k$-th bit of the value of $v$ in the solution vector ${\bf v}$ is $1$ or $0$. Since $\mathcal{T}$ runs in poly-logarithmic space, $\mathcal{T}$ would have a read-only input tape that explicitly describes the system of equations, the variable $v$, and the index $k$, alongside an additional separate work tape that uses $O(\log^2 n)$ space~\cite{sipser2006}.

Since we are interested in the hitting probabilities of $\mathbb{G} \times \pi$, we denote the system of equations that solve for these hitting probabilities as $S_{\mathbb{G},\pi}$. While our system of equations $S_{\mathbb{G},\pi}$ is guaranteed a unique solution by Lemma~\ref{uniquesolution}, $S_{\mathbb{G},\pi}$ can only be implicitly constructed, since constructing every equation would lead to an exponential blow-up, as $\mathbb{G} \times \pi$ is of exponential size w.r.t. the input of $\mathbb{G}$ and $\pi$. Now imagine using $\mathcal{T}$ to solve $S_{\mathbb{G},\pi}$. It would need $S_{\mathbb{G},\pi}$ represented explicitly on its input tape, but, once given, would only use polynomial space on its work tape as desired. Therefore, the problem becomes how to represent the input tape without constructing it explicitly.

Note that all the parameters in $S_{\mathbb{G},\pi}$ can be computed in PSPACE from $\mathbb{G}$ and $\pi$, as they are either the elements of the transition matrix $P$, as shown in Lemma~\ref{transprobs}, or the constant $1$. Thus, we can ``virtualize" the read-only input tape containing the exponentially large $S_{\mathbb{G},\pi}$. We can simulate read/move operations of $\mathcal{T}$ on its input tape using polynomial space to record the position of the reading head and using polynomial space to simulate reading from the input by computing these values from $\mathbb{G}$ and $\pi$.   When applied to $S_{\mathbb{G},\pi}$, the end result is a polynomial-space Turing machine that takes a state $r \in \mathbb{G} \times \pi$ and an index $k$ as input and returns the $k$-th bit of $h_r$.

\begin{thm}
    Given a state $r \in \mathbb{G} \times \pi$, it is possible to compute the value of the $k$-th bit of $h_r$,  the probability that the set $T$ is reached from $r$, in PSPACE.
\end{thm}

\subsection{Strategy Improvements}\label{improve}

In the previous section, we showed that given a history $h$ and an Agent $i$ it was possible to determine the probability that Agent $i$ reached his goal set $G_i$ in $\mathbb{G}|_h$ when all agents followed $\pi|_h$ in PSPACE by analyzing the Markov Chain $\mathbb{G} \times \pi$. 

By definition, we know that if $\pi$ is not a subgame-perfect equilibrium, then there exists some history $h$ such that Agent $i$ has a profitable deviation from $\pi|_h$ in $\mathbb{G}|_h$. We do not know, however, what the strategy that realizes this deviation looks like and if it can be represented in polynomial space. In this section, we address these questions. We also introduce the notation of the \emph{value} of a state to refer to the value of its hitting probability 
 (see Lemma~\ref{uniquesolution} above), 
 following common texts in Markov Chains such as~\cite{Puterman94}.

First, we construct the Markov Decision Process (MDP) $\mathbb{G}_i \times \pi$. Since the idea behind this MDP is very intuitive, whereas its formal specification is cumbersome, we opt for an informal explanation of its construction. The MDP $\mathbb{G}_i \times \pi$ has the same state space of $V \times S \times [F]$ as the Markov Chain $\mathbb{G} \times \pi$. But at each state $v \in V \times S \times [F]$, Agent $i$ can choose an arbitrary distribution $d \in \mathbb{D}(A_i)$ over $A_i$. Note that this distribution may be different from the distribution that the original strategy $\pi_i$ would output. The transition probabilities between the states then work exactly as in $\mathbb{G} \times \pi$ if the output of $\pi_i$ were $d$ instead, keeping all the other outputs of $\pi_j$ for $j \not = i$ unchanged. This allows us to compute the transition probabilities in PSPACE, as in Lemma~\ref{transprobs}.

The idea behind $\mathbb{G}_i \times \pi$ is to analyze the payoffs available to Agent $i$ when he is allowed to vary his strategy and deviate from $\pi_i$. Nevertheless, if we change Agent $i$'s output at every state $v \in V \times S \times [F]$, then we end up creating an exponentially large new strategy, placing us outside of PSPACE. The idea in this section is to prove the existence of a state $v^*$ such that, after only a ``small, local" change to the output of Agent $i$ at $v^*$, the value of $v^*$ is immediately improved. Formally,  this means that there is a relevant history-reachable state $v^* \in R$ such that the strategy $\pi_i$ needs only to modify its output at $v^*$ to some deterministic distribution, i.e., a deterministic choice of action, in order to strictly increase the hitting probability $h_{v^*}$ of reaching $T$ from $v^*$. Connecting this back to $\mathbb{G}$, this means that only a very simple change needs to be made to $\pi_i$ in order to show that it is not optimal after the history $h$ corresponding to $v^* \in \mathbb{G} \times \pi$ is observed in $\mathbb{G}$.

\begin{lem}\label{existenceimprovement}
    If $\pi$ is not a subgame-perfect equilibrium in $\mathbb{G}$, then there exists a state $v^*$ in $\mathbb{G} \times \pi$ such that (1) $v^* \in R$ and (2) the probability of reaching the target set $T$ from $v^*$ can be strictly improved by changing the output of $\pi_i$ at $v^*$ to a deterministic distribution over $A_i$. 
\end{lem}

\begin{proof}

    Since $\pi$ is not a subgame-perfect equilibrium in $\mathbb{G}$, it must be the case that after witnessing some history $h$, the value of the state $r \in R$ resulting from the history $h$ in $\mathbb{G} \times \pi$ is not optimal. If this were not the case, the value would be optimal in all subgames, and Agent $i$ would not have a profitable deviation, contradicting the assumption that $\pi$ is not a subgame-perfect equilibrium. In order to find this state $r$, we vary Agent $i$'s actions in the MDP $\mathbb{G}_i \times \pi$.
    
    We now work by backwards induction. Consider the subset of $R$ given by $\langle v, s, n | n = F \rangle$. Clearly, all choices of actions, including the one suggested by $\pi_i$, are optimal as these states have no outgoing transitions. Now consider the subset of $R$  given by $\langle v, s, n | n = F - 1 \rangle$. Since the value at each of the states in $\langle v, s, n | n = F \rangle$ is optimal by default, it can be determined if $\pi_i$ suggests an optimal distribution by maximizing over all actions in $A_i$ -- $\mathbb{G}_i \times \pi$ is an MDP, so at least one of the distributions that achieve an optimal value is deterministic~\cite{Puterman94}. If all such values are indeed optimal, then we continue to the subset of $R$ given by $\langle v, s, n | n = F - 2 \rangle$. Eventually, we must reach a state $v$ where the value is not optimal. Let $v^* = \langle v, s , n^* \rangle$ be the first such state observed; if there are multiple states with the same largest time index $n^*$, then we sort them lexicographically. We are not interested in solving $\mathbb{G}_i \times \pi$ explicitly; this process terminates after a single change has been made at one state $v^*$.
\end{proof}

This proof guarantees the existence of a state $v^*$ with the desired properties of Lemma~\ref{existenceimprovement}, if $\pi$ is not a subgame-perfect equilibrium.  We can then create a new Markov Chain $\mathbb{G} \times \pi'$ (representing that the strategy $\pi$ has changed slightly) that only changes the outgoing transition probabilities at $v^*$ by considering the deterministic distribution found in Lemma~\ref{existenceimprovement}. The new probability that the target set $T$ is reached from $v^*$ in $\mathbb{G} \times \pi'$ as opposed to $\mathbb{G} \times \pi$ can then be computed in the same manner as Section~\ref{computation}. 

\subsection{Comparing the Two Probabilities}\label{compare}

In Section~\ref{improve}, we saw that if a strategy profile $\pi$ is not a subgame-perfect equilibrium in a game $\mathbb{G}$, then there exists a reachable state $v^*$ in $\mathbb{G}_i \times \pi$ where an Agent $i$ can improve his value at $v^*$ by switching his action choice at $v^*$ to a deterministic distribution. This means that for some history $h$ and deterministic distribution to play at $v^*$, which we can guess in NPSPACE=PSPACE, the value of $v^*$ in $\mathbb{G} \times \pi'$ is strictly greater than the one in $\mathbb{G} \times \pi$ if $\pi$ is not a subgame-perfect equilibrium in $\mathbb{G}$. As shown in Section~\ref{computation}, we have a PSPACE algorithm to query the bits of the hitting probabilities of $v^*$ in both $\mathbb{G} \times \pi$ and $\mathbb{G} \times \pi'$.  

We can compare these probabilities through a lexicographic comparison in which we query and compare the individual bits of both probabilities and use a counter to keep track of the compared index. The final issue is then to determine if this counter can always be represented in polynomial space. In order to do this, we use backwards induction to reason about the maximum size of the representation of a hitting probability. 

\begin{lem}\label{sizeofprobs}
    All hitting probabilities in both $\mathbb{G} \times \pi$ and $\mathbb{G} \times \pi'$ have a representation that is at most exponential in the size of $\mathbb{G}$ and $\pi$.
\end{lem}

\begin{proof}
The arguments for $\mathbb{G} \times \pi$ and $\mathbb{G} \times \pi'$ are the same, so WLOG, we work with $\mathbb{G} \times \pi$. Our proof uses backwards induction, so we start at the subset of states $\langle v , s ,n | n = F \rangle$.  For a state $x \in \langle v , s ,n | n = F \rangle$, the hitting probability $h_x$ is either $0$ or $1$ depending on whether it is a target state or not. The intuitive idea behind this proof is to define subsets $\langle v , s ,n | n = k \rangle$ for $k \in [F]$ and show that when moving backwards through the index $n$, the length of the representation for the hitting probabilities grows \emph{additively}. Since there are exponentially many such subsets (since $|F|$ is exponential), we end up \emph{adding} exponentially many times, giving us an exponential-sized result.  

Now, consider the set of states $\langle v , s ,n | n = F - 1 \rangle$. For a state $x$ in $\langle v , s ,n | n = F - 1 \rangle$ the hitting probability $h_x$ can be calculated by the equation $h_x = p_a \cdot h_a + p_b \cdot h_b \ldots $ where the hitting probabilities $h_a$ and $h_b$ on the right-hand side (RHS) refer to states in the set  $\langle v , s ,n | n = F \rangle$.  Recall the definitions of $b$, the concurrency bound, $\mathbb{L}$, the size bound of probabilities, and $A$, the set of actions from Definition~\ref{pcgs} and the definition of $\mathbb{G} \times \pi$ from Section~\ref{subgameperfect}.  We know the following facts about the RHS :
\begin{enumerate}

 \item The number of terms on the RHS has an upper bound of $|V| \cdot |S|$ (recall that $S$ is the state space of the product transducer that represents $\pi$), an upper bound on the size of the subset of the state space of $\mathbb{G} \times \pi$ with a fixed time index.
        This upper bound is singly exponential in the size of the input of $\mathbb{G}$ and $\pi$ due to the presence of $S$.
     
        \item The probabilities $p_i$ of transitioning from the state $\langle v_1, s_1, n_1 \rangle$ to $\langle v_2, s_2, n_2 \rangle$ are either $0$ or given by the expression $ \sum_{\theta \in supp(O(s_1,v_1))} \mathbb{P} ( O(s_1,v_1)[\theta]) \cdot \delta(v_1, \theta, v_2)$, as shown at the start of Section~\ref{subgameperfect}. The term $\mathbb{P} ( O(s_1,v_1)[\theta])$ refers to the probability that the product transducer $T_{\pi}$, as given in Definition~\ref{product},  assigns to  $\theta$ on an input of $v_1$ in state $s_1$.  This, itself, is the product of at most $b$ probabilities from the individual transducers $\pi_i$. Each probability in each transducer $\pi_i$ is exactly $\mathbb{L}$ bits in length. Therefore, $\mathbb{P} ( O(s_1,v_1)[\theta])$ is at most $b \cdot \mathbb{L}$ bits in length. Furthermore, we multiply this probability by $\delta(v_1, \theta, v_2)$, which is also $\mathbb{L}$ bits in length, giving us a total of $(b+1) \cdot \mathbb{L}$ bits. Finally, we have a sum over $supp(O(s_1,v_1))$, which has an upper bound of $|A|^{b}$ elements.  Thus, the total number of bits used to represent a transition probability $p_i$  in $\mathbb{G} \times \pi$ is $\log(|A|^b) + (b+1) \cdot \mathbb{L}$.  We refer to this number as $\mathbb{L}_{\mathbb{G} \times \pi }$, the size of the representation of the probabilities in $\mathbb{G} \times \pi$. As we have already shown in Lemma~\ref{transprobs}, $\mathbb{L}_{\mathbb{G} \times \pi}$ is polynomial in the size of the input.

\end{enumerate}
Let us consider the equation $h_x = p_a \cdot h_a + p_b \cdot h_b \ldots $. We claim that the size of the representation of the evaluation of the RHS has an upper bound of $\log(|V| \cdot |S|) + \mathbb{L}_{\mathbb{G} \times \pi }$. Each term $h_i \cdot p_i$ has  a representation of size $\mathbb{L}_{\mathbb{G} \times \pi }$, since $h_i$ is either $0$ or $1$. Also, since there are at most $|V| \cdot |S|$ terms on the RHS, we are at most considering a sum over $|V| \cdot |S|$ elements, which can be upper bounded by multiplying by  $|V| \cdot |S|$, adding at most $\log(|V| \cdot |S|)$ bits.  Therefore, it is possible to represent the hitting probability $h_x$ for a state $x \in \langle v , s ,n | n = F - 1 \rangle$ using at most $\log(|V| \cdot |S|) + \mathbb{L}_{\mathbb{G} \times \pi }$ bits.

An induction argument shows that for the set $\langle v, s,n | n = F - c \rangle$, we have a hitting probability representation bound of $c(\log(|V| \cdot |S|) + \mathbb{L}_{\mathbb{G} \times \pi })$. As we have shown in the previous paragraph, the bound holds for the set $\langle v, s,n | n = F - 1 \rangle$.  Now, assume that the bound $(c-1)(\log(|V| \cdot |S|) + \mathbb{L}_{\mathbb{G} \times \pi })$ holds for the set $\langle v, s,n | n = F - (c-1) \rangle$ and set up the equation $h_x = p_a \cdot h_a + p_b \cdot h_b \ldots $ again for $x \in \langle v, s,n | n = F - c \rangle$. The terms $p_i \cdot h_i$ on the RHS have a representation of size $\mathbb{L}_{\mathbb{G} \times \pi} + (c-1)\log(|V| \cdot |S| ) + (c-1)\mathbb{L}_{\mathbb{G} \times \pi }$, corresponding to multiplying the $\mathbb{L}_{\mathbb{G} \times \pi}$ length probability $p_i$ by the $(c-1)(\log(|V| \cdot |S|) +\mathbb{L}_{\mathbb{G} \times \pi })$ length hitting probability $h_i$ (our inductive assumption since $i \in \langle v, s,n | n = F - (c-1) \rangle$).  There are at most $|V| \cdot |S|$ such terms on the RHS, corresponding to multiplying by $|V| \cdot |S|$, which adds $\log(|V| \cdot |S| ) $ bits. In total, we have $\log(|V| \cdot |S| ) + \mathbb{L}_{\mathbb{G} \times \pi} + (c-1)\log(|V| \cdot |S|) + (c-1)\mathbb{L}_{\mathbb{G} \times \pi } = c(\log(|V| \cdot |S|) + \mathbb{L}_{\mathbb{G} \times \pi })$

Finally, note that this parameter $c$ can never exceed $|F|$, as after $|F|$ steps, we reach the initial state that corresponds to time step $1$. Therefore, all hitting probabilities have a representation of size at most   $|F|( \log(|V| \cdot |S| ) + \mathbb{L}_{\mathbb{G} \times \pi })$. This upper bound is singly exponential in the size of the input due to the presence of $|F|$.
\end{proof}

By Lemma~\ref{sizeofprobs}, the hitting probabilities of $v^*$ in $\mathbb{G} \times \pi$ and $\mathbb{G} \times \pi'$ have a representation that is at most singly exponential in the size of the input.  Therefore, the counter that keeps track of the bit index we are comparing has at most polynomial size. This is the final component needed for our PSPACE upper bound. 

\begin{thm}\label{upper}
The problem of verifying whether a given strategy profile $\pi$, represented by a tuple of
probabilistic transducers, is a subgame-perfect equilibrium in a given $b$-bounded probabilistic concurrent game  $\mathbb{G}$ can be decided in PSPACE. 
\end{thm}

\subsection{Alternate Models and Transducers}\label{alternative}

\subsubsection{Motivation Behind Deterministic Transition Functions in Strategy Transducers} In Definition~\ref{probabilistictransducermodel}, we introduced finite-state transducers with deterministic transition functions as models of strategies. In this section, we elaborate on the motivation behind this choice, which boils down to two main points -- these transducers simplify our presentation and are mathematically sufficient models. Using a deterministic transition function, we guaranteed that after reading a history $h \in V^+$, a transducer $\pi_i$ would be a single state $s \in S_i$. This allowed us to directly connect subgames to individual states in $\mathbb{G} \times \pi$, simplifying the presentation of Section~\ref{subgameperfect}. We also note that it can be shown that Nash equilibria and subgame-perfect equilibria always exist in our concurrent game model through the use of backward induction. Backwards induction would result in \emph{Markovian} strategies for each agent, i.e., strategies of type $V \times [F] \rightarrow \mathbb{D}(A_i)$~\cite{mpe}. Therefore, considering transducers with deterministic transitions that model functions of type $V^+ \rightarrow \mathbb{D}(A_i)$ subsumes the family of Markovian strategies (the value corresponding to $[F]$ can be obtained from the length of an element $h \in V^+$) and is therefore sufficient -- there is no need to consider more general models, since both equilibria always exist when the model presented is considered.

\subsubsection{Full Concurrency Model}
In Section~\ref{subgameperfect}, we constructed a Markov Chain $\mathbb{G} \times \pi$ with the critical property that elements of its transition table $P$ could be queried on-the-fly in PSPACE. We now show that the elements of $P$ -- representing the probability that a state $\langle v_1,s_1,n_1 \rangle$ transitions to a state $\langle v_2, s_2, n_2 \rangle$ --  could be computed in PSPACE even when a full concurrency model is considered as opposed to the $b$-bounded one presented in this paper.  Recall that elements of $P$ were given by a product $p_v \cdot p_s \cdot p_n$. In a full concurrency model, $p_s$ would not change since it considers every strategy $\pi_i$ to begin with. Additionally, $p_n$ would not change as well since it only accounted for the time index. The only change would be to $p_v$, which was calculated through the sum $ \sum_{\theta \in supp(O(s_1,v_1))} \mathbb{P} ( O(s_1,v_1)[\theta]) \cdot \delta(v_1, \theta, v_2)$. This sum ranged over   $ supp(O(s_1,v_1))$, which could only contain polynomially many elements w.r.t the input due to the $b$-bounded nature of the game. Even if, however, it ranged over $\Theta = \bigtimes_{i \in \Omega} A_i$, the cross-product of every action set as in a fully concurrent setting, this would still only be an exponential-size set. It could, therefore, be iterated over in PSPACE, meaning that $p_v$ could still be computed in PSPACE. The rest of the results would then follow in exactly the same manner, giving us an analogous statement of Theorem~\ref{upper} for the subgame-perfect equilibrium verification problem in full concurrency models. Similarly, the EXPTIME upper bound presented in Theorem~\ref{nashupper} for the Nash equilibrium verification problem would also be preserved in full concurrency models. 

\subsubsection{Probabilistic Transition Functions}
We now consider what would happen if we were to consider a transducer model with a probabilistic transition function as opposed to a deterministic one. In $\mathbb{G} \times \pi$ as defined in Section~\ref{subgameperfect}, the probability that a state $\langle v_1,s_1,n_1 \rangle$ transitions to a state $\langle v_2, s_2, n_2 \rangle$ could still be computed in PSPACE. Recall that this probability was given as the product of three terms, $p_v \cdot p_s \cdot p_n$. The only change to this would be to $p_s$, the transition probability between $s_1 = \langle s_1^1 \ldots s_1^k \rangle$ and   $s_2 = \langle s_2^1 \ldots s_2^k \rangle$. This probability would then be given by the product $ \bigtimes_{i \in \Omega} \mathbb{P}(\gamma^i(s^i_1,v_1)[s^{i}_2])$, where $\mathbb{P}(\gamma^i(s^i_1,v_1)[s^{i}_2])$ is the probability the distribution $\gamma^i(s^i_1,v_1)$ assigns to $s^i_2$. This product could be computed in PSPACE as well; if $m$ is the size of the largest encoding of a probability in $\pi$, then the encoding of the product would be at most $\log(2^{m\cdot k})$ bits, which is polynomial w.r.t the input.

However, for probabilistic transition functions, the question now becomes whether an agent should be able to view the current state of the other transducers or not in $\mathbb{G} \times \pi$. If knowledge of the other transducers includes their current state, the analysis proceeds exactly as in Section~\ref{subgameperfect}, as we would again have a correspondence between states of $\mathbb{G} \times \pi$ and subgames induced by histories. If this is not the case, then following a history $h \in V^+$ would correspond to a distribution over states in $\mathbb{G} \times \pi$, accounting for the various states that each transducer $\pi_i$ could be in after reading $h$. Given that we feel this scenario corresponds less directly to the intention of the definition of Nash equilibrium and that our transducer model sufficiently captures the set of Markovian strategies, we leave this question for future work.

\section{Nash Equilibrium}

In this section, we demonstrate that the Nash equilibrium verification problem is EXPTIME-complete, starting with the upper bound.

\subsection{Upper Bound}
\begin{thm}\label{nashupper}
    The problem of verifying whether a given strategy profile $\pi$, represented by a tuple of probabilistic transducers, is a Nash equilibrium in a given $b$-bounded probabilistic concurrent game $\mathbb{G}$ can be decided in EXPTIME.
\end{thm}

\begin{proof}
    The algorithm to decide the Nash equilibrium verification problem proceeds in two steps. For each Agent $i$, it first calculates the payoff that Agent $i$ receives under $\pi$ (following $\pi_i$ without deviating). This can be rephrased as the hitting probability of the set $T = \langle v,s,n | v \in G_i \rangle$ from the initial state in $\mathbb{G} \times \pi$, as constructed in Section~\ref{subgameperfect}. As shown in Section~\ref{computation}, this hitting probability can be computed in PSPACE.

    The next step is to compute the probability of the payoff achieved by the most profitable deviation from Agent $i$. This can be done by explicitly constructing and solving the MDP $\mathbb{G}_i \times \pi$ described in Section~\ref{improve} (a step we explicitly avoided in Section~\ref{subgameperfect}). In every Markov Decision process, there is a policy (a deterministic, memoryless strategy) that achieves the optimal value~\cite{Puterman94} from every state simultaneously. We can solve for this policy through backwards induction~\cite{Puterman94}, and since this solution is guaranteed to return deterministic distributions at each state, there is no issue of representation when it comes to the numerical values used in the policy. Once this optimal policy is fixed, applying Lemma~\ref{sizeofprobs} gives us an exponential bound (in the size of the original input of $\mathbb{G}$ and $\pi$) on the size of the representation of the optimal hitting probability from the initial state.
    
    Explicitly represented MDPs can be solved in polynomial time~\cite{MDPisPc,Puterman94}. Therefore, scaling these polynomial-time algorithms up to the explicitly constructed exponentially large $\mathbb{G}_i \times \pi$ gives us an EXPTIME upper bound for computing the optimal hitting probability from the initial state.   
    Once both hitting probabilities have been computed —- the original and the computed optimal —- the only remaining step is to compare them. This can be done in EXPTIME since the numbers involved have at most exponential length w.r.t the input.  If the optimal value from the initial state in $\mathbb{G}_i \times \pi$ is larger than the value from the initial state in $\mathbb{G} \times \pi$, then Agent $i$ has a profitable deviation, and $\pi$ is not a Nash equilibrium.  By checking each agent in this way, we can decide whether some agent has a profitable deviation in EXPTIME and, therefore, decide the Nash equilibrium verification problem in EXPTIME.
\end{proof}
Broadly speaking, an EXPTIME upper bound for the verification problem appears in various other works in various other forms as a ``folk theorem," e.g.~\cite{backwardsinduction}. In this work, we make the details explicit and consider issues of numerical representation that are often ignored in other works.

\subsection{Lower Bound}
In order to demonstrate our lower bound reduction, we recall the definition of an \emph{alternating} Turing machine.  
\begin{defi}[Alternating Turing Machine]
An alternating Turing Machine $M$ is a $5$-tuple $M = \langle R, \Gamma , r_0, \Delta,\kappa \rangle$ with the following interpretations : 
\begin{enumerate}
    
\item  $R$ is the set of states. $r_0 \in R$ is the initial state. 
    \item $\Gamma$ is the alphabet. WLOG, we assume it contains an empty symbol $\emptyset$.
    \item $\Delta : R \times \Gamma \rightarrow \mathcal{P}(R \times \Gamma \times \{\rightarrow,\leftarrow \} )$ 
        is the transition function, where $\mathcal{P}(\cdot)$ refers to the powerset construction. The set $\{\rightarrow,\leftarrow \}$ represents the left and right directions -- how the head moves over the tape in the Turing machine. Since this is an alternating Turing machine, the range of the transition function is a powerset representing all possible transitions available from a single state-character pair. For notational convenience, we assume that the range of $\Delta$ on an arbitrary input has at most size two, i.e., there are only two possible successors from each state-character pair. This assumption does not change the complexity-theoretic properties of the decision problem from which we wish to reduce.
    \item $\kappa: R \rightarrow \{ \mathrm{accept}, \mathrm{reject}, \vee, \wedge, -\}$ is the state labeling function that associates each state with its type. A state can be accepting, rejecting, existential ($\vee$), universal ($\wedge$), or deterministic ($-$).

\end{enumerate}
\end{defi}

Alternating Turing machines are a generalization of non-deterministic Turing machines. In a non-deterministic Turing machine, it is possible for a machine state $r$ to transition to multiple possible successors. Without loss of generality, we can assume that there are at most two possible successors, which we call the $\alpha$ successor $r_{\alpha}$ and the $\beta$ successor $r_{\beta}$.  The computation from state $r$ is accepted if the computation is accepted starting from at least one of the successors. Informally, in non-deterministic states where multiple choices are possible, the machine can ``choose" the best option and proceed from there.

Alternating Turing machines generalize non-deterministic Turing machines by classifying machine states as existential ($\vee$), universal ($\wedge$), or deterministic ($-$). Existential states work exactly as described previously in the description of non-deterministic Turing machines. In an existential state, multiple transitions are possible, and the machine can choose the one that best serves the purpose of acceptance. The dual of this concept is the universal state, in which multiple transitions are possible, but instead of choosing one, the computation must be accepting from \emph{all} successor states. Deterministic transitions work as one would expect from a deterministic Turing machine.  While it is possible to encapsulate deterministic behavior by creating either an alternating or universal state with a single successor, it is more convenient notationally to introduce a special category for deterministic states.

The presence of universal states means that the computations in an alternating Turing machine are represented as \emph{trees} instead of paths. Since universal states require that both successors accept, such transitions are precisely the areas where the tree branches. For more precise details, we refer the reader to~\cite{ChandraStockmeyerAlternation}.

\subsubsection{A High Level View of the Reduction}

First, we introduce the problem we aim to reduce from.
\begin{problem*}    
Given an alternating Turing Machine $M$ and a natural number $n \in \mathbb{N}$ written in unary, does $M$ accept the empty tape using at most $n$ cells? 
\end{problem*}

This problem is hard for the class APSPACE = EXPTIME~\cite{ChandraStockmeyerAlternation}.
In this decision problem, the machine $M$ can never use more than $n$ cells in an accepting computation, as any computation that tries to access more than that amount of space is immediately rejected.
Since the machine can never use more than $n$ cells, we introduce the concept of an ID -- a string that represents the complete information of the tape and machine at a given moment in time. 
\begin{exa}For example, an ID may read $122\langle r_3, 5 \rangle68$.  This string conveys the following information: {\bf (1)} The string is $6$ characters long -- $\langle r_3,5 \rangle$ is considered to be a single character. This means that the bounded Turing machine can only use $6$ cells, i.e., $n=6$.
{\bf (2)} The first cell, counting from the left, reads $`1'$. For this example, $\Gamma$ is given by the set of digits.
{\bf (3)} The head of the Turing machine is currently reading the fourth cell from the left, represented by the character $\langle r_3,5 \rangle$. The machine is in state $r_3 \in R$, reading the character $'5'$ written on the fourth cell from the left.

\end{exa}
An ID contains all the necessary information to compute the possible next stages of computation when paired with the transition function $\Delta$.  WLOG, we assume the head of the machine starts positioned at the leftmost of these cells, meaning the initial ID of this $6$-cell machine is given by $ \langle r_0, \emptyset \rangle \emptyset \emptyset \emptyset \emptyset \emptyset $. This ID corresponds to an empty $6$-cell tape with the head in $r_0$ at the leftmost cell. Since the ID is of length $n$, the number of possible IDs of the machine $M$ is exponential in $n$.

Due to the complexity of the reduction, we first present a very high-level overview of the main ideas and concepts used before moving on to the technical details, which we interleave with further intuition. The ultimate goal of this reduction is to create a $1$-bounded concurrent game $\mathbb{G}$ with $n+1$ agents and a strategy profile $\pi = \langle \pi_1 \ldots \pi_{n+1} \rangle$ such that $\pi$ is a Nash equilibrium in $\mathbb{G}$ iff the machine $M$ does {\bf not} accept the empty tape using at most $n$ cells in its computation.  This construction uses the first $n$ agents and their strategies to ``simulate" $M$. We leave Agent $n+1$ for last in the discussion, as their role is the most complicated, and their inclusion constitutes the most novel part of this reduction. The first step is to describe how the machine $M$ can be simulated using $\mathbb{G}$ and $\pi$.

The high-level idea behind the simulation of $M$ is to track its computations by tracking its IDs. The number of IDs of $M$ is, however, exponential in the size of $M$. Therefore, we opt to store them in a \emph{distributed} manner. The verification problem we construct contains $n$ agents that correspond to the $n$ characters of $M$'s ID strings. The $i$-th character of the ID is stored in the state of $\pi_i$.  These $n$ agents have a reachability goal of the entire set $V$, the set of states of the game $\mathbb{G}$. Therefore, they always receive a constant payoff of $1$ regardless of what strategies are employed and have no incentive to deviate from the strategies assigned to them by $\pi$. As shorthand, we refer to these agents as ``simulators'', since their only purpose is to simulate the computations of $M$ by tracking its IDs. The general idea behind the representation of the IDs of $M$ is depicted in Figure~\ref{fig:simulation}.

In order to update the $i$-th character in the current ID, Agent $i$ needs global information about the transitions in $M$. This information is stored in the states in $V$. Intuitively, when the head of $M$ is positioned over cell $i$, the strategy $\pi_i$ outputs an action that describes the complete information of the transition -- which character was written, which cell the head was just at (in this case, cell $i$), which state the machine has transitioned to, and which direction the head is moving in. Once Agent $i$ outputs this action, the game $\mathbb{G}$ deterministically transitions to a state that is labeled by this same information, allowing all agents to view and update the internal state of their strategy representing their ID character. For example, if Agent $i-1$ sees that the head was previously at cell $i$ and is moving left in state $q$, they know the head will be positioned over their cell in state $q$ next and can therefore update the state of $\pi_{i-1}$ accordingly. The state of $V$ then deterministically transitions to a state indicating that the head is over cell $i-1$ and it is awaiting an action from Agent $i-1$.

While this scheme works for deterministic transitions, the alternating Turing machine $M$ has universal and existential transitions as well.  The high-level idea behind universal transitions is to output each transition with $.5$ probability. Therefore, in order for the simulation of the machine to reach an accepting ID configuration with probability $1$, it must be the case that $M$ has an accepting computation tree, e.g., a computation tree where every branch is accepting. The high-level idea behind existential transitions is that Agent $n+1$ is allowed to choose the existential transitions. If the machine $M$ is in an existential state over the cell $i$, it allows Agent $n+1$ to choose which of the two possible successors should be taken. This information is once again encoded into the state space $V$, and Agent $i$ uses this new information when $\pi_i$ outputs their transition action. 

This brings us to the last agent, Agent $n+1$. Agent $n+1$ has two roles in the game. The first role is to decide how the existential transitions of $M$ are resolved. Agent $n+1$ is incentivized to choose existential transitions of $M$ that lead to acceptance, since their reachability goal is a single state that is visited after the machine $M$ reaches an accepting state. There is, however, another way to reach this state, which brings us to their second role. Intuitively, Agent $n+1$ chooses whether to ``start" the simulation of $M$ at the very first time step. If they choose to start the simulation, then they can only reach their goal if an accepting ID of $M$ is reached. Since the universal transitions are resolved probabilistically, a non-accepting $M$ means that this state is reached probabilistically, and an accepting $M$ means this state is reached with probability $1$. 

Alternatively, Agent $n+1$ can choose not to start the simulation of $M$ and instead take a different transition that offers them a fixed payoff that is independent of whether $M$ accepts or not. In fact, this is the choice that the input strategy $\pi_{n+1}$ makes. This payoff is designed to be higher than the payoff that can be achieved by all non-accepting machines $M$, yet still lower than $1$. Therefore, Agent $n+1$ is only incentivized to deviate from $\pi_{n+1}$ iff they can choose the existential transitions of $M$ to create an accepting computation tree for $M$ and therefore reach their goal with probability $1$.

\begin{figure}[ht]
  \centering
  \includegraphics[width=\textwidth]{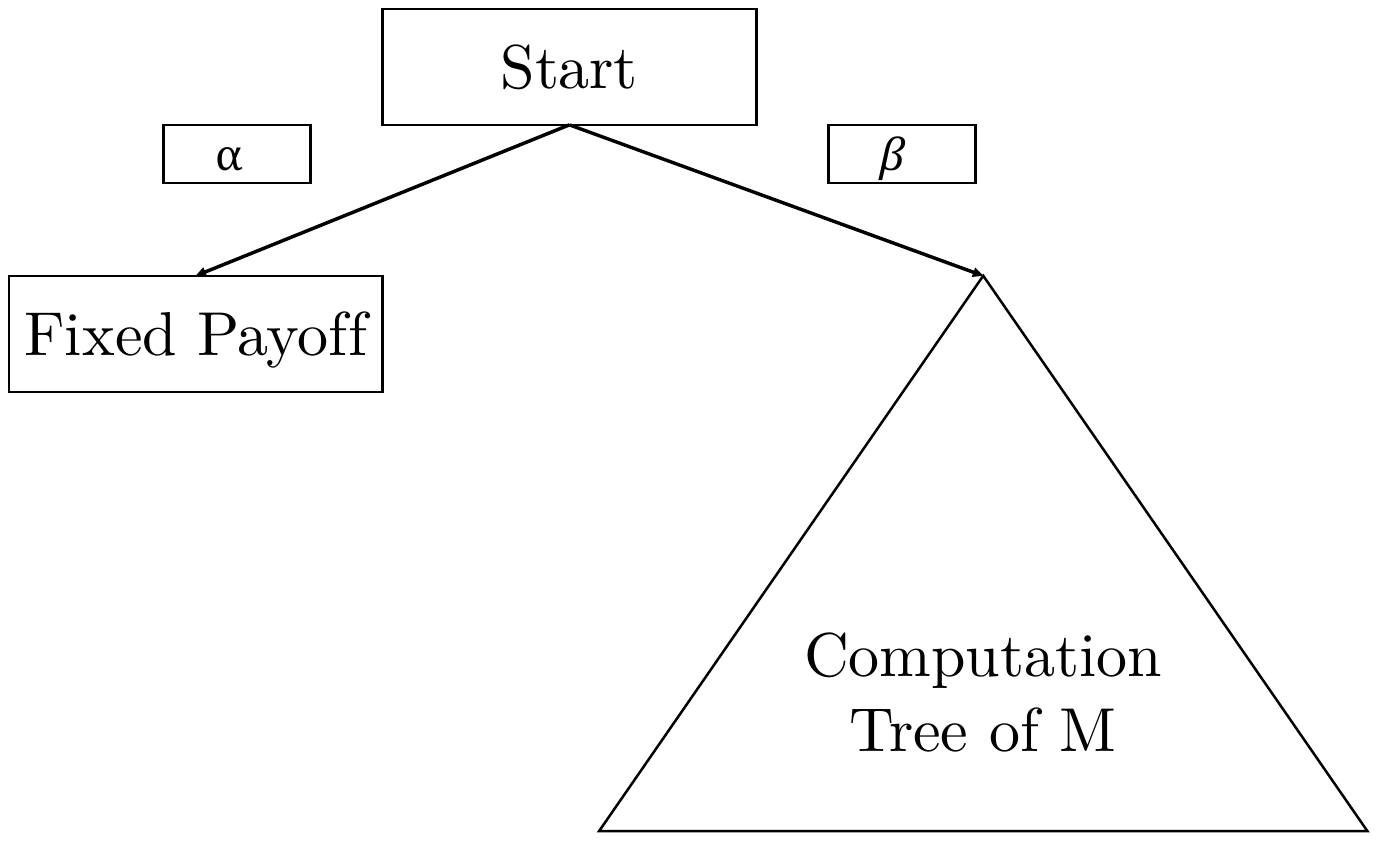}
  \caption{The crucial mechanism of the reduction, Agent $n+1$'s first choice. The Agent can either choose the fixed payoff mechanism with $\alpha$, or start the probabilistic simulation of $M$ by the other $n$ agents with $\beta$, with the intention of reaching an accepting configuration. Since $M$ is an alternating Turing machine, it admits a computation tree. If Agent $n+1$ chooses $\beta$, the game proceeds by probabilistically sampling one of the branches in the computation tree. The numbers are chosen in such a way that if even one branch in the computation tree is nonaccepting (meaning $M$ is nonaccepting), Agent $n+1$ gets a better chance of reaching their goal through choosing $\alpha$ and taking the fixed payoff mechanism.}
  \label{fig:reductionidea}
\end{figure}

Before moving on to the technical details, we provide a brief overview of the reduction idea. The strategy $\pi$ tells Agent $n+1$ to take the fixed payoff at the start of the game. Therefore, the machine $M$ is not being simulated, and Agent $n+1$ is not deciding between the $\alpha$ and $\beta$ successors in the simulation of $M$. This fixed payoff is chosen to be higher than the payoff obtained by any non-accepting $M$, so if $M$ does not accept, then $\pi$ is a Nash equilibrium. If, however, the machine $M$ has an accepting computation tree, then there is a strategy for Agent $n+1$ to simulate the machine $M$ and receive a payoff of $1$, meaning that $\pi$ is not a Nash equilibrium. This is the central idea behind the reduction, and it is illustrated in Figure~\ref{fig:reductionidea}.

\subsubsection{Game Construction}\label{gameconstruction}
In this section, we provide details on how to construct the $1$-bounded probabilistic concurrent game (i.e. a turn-based game) $\mathbb{G} = \langle V, v_0, \Omega, A = \{A_1,A_2, \ldots A_{n+1}\},P, \delta,  \mathbb{L}, G = \{G_1, G_2, \ldots G_{n+1}\} , F \rangle$.

There are two high-level ideas behind this construction. The first is to use the state space of $\mathbb{G}$ to relay information about the transitions of $M$ to each agent in the game. The second is to construct the mechanism that creates the ``fixed payoff" that Agent $n+1$ receives under $\pi$. 
\begin{enumerate}
    \item The state set $V = V_1 \cup V_2 \cup V_3 \cup \{ { \sf start}, { \sf base}, { \sf goal}, {\sf sink} \}$ consists of several components, all of which have polynomial size with respect to     $|M|+n$.
    \begin{enumerate}
        \item The set $V_1$ is $[n] \times \{\emptyset, \vee\}$, where $[n]$ denote $\{1,2 \ldots n\}$. These states keep track of the position of the head and whether the current state of the machine is existential or not. Recall that when $M$ is an existential state over cell $i$, Agent $i$ needs to first be informed about which successor Agent $n+1$ wants before outputting the transition information. 
        \item The set $V_2$ is $[n] \times \Gamma \times \{\rightarrow,\leftarrow \} \times R$ . These states correspond to the information made by a transition -- the last position in $[n]$ of the head, the character in $\Gamma$ written by the head, the direction $\rightarrow$ or $\leftarrow$ the head is moving in, and the new state in $R$ of the machine. When a state $v \in V_2$ is visited, this symbolizes that a transition has been made in $M$. The state $v$ then contains all the necessary information for each agent in the game to accurately track the ID of $M$ through the state spaces of the individual $\pi_i$ transducers. 
        \item The set $V_3$ is $[n] \times \{\alpha,\beta \}$. 
        These states are used to record the existential choices of Agent $n+1$. If the game was previously in a state $\langle i, \vee \rangle \in V_1$, then Agent $n+1$ may specify which existential successor ($\alpha$ or $\beta$) they would like to see. Once this is specified, the game transitions to either $\langle i , \alpha \rangle$ or $\langle i,  \beta \rangle \in V_3$. This represents the aforementioned intermediate step used to deal with existential states used to relay the choices of Agent $n+1$ to the other agents. 
        \item Finally, there are four additional states labeled ${\sf start}$, ${\sf base}$, ${\sf goal}$, and ${\sf sink}$. These states are used to create the fixed payoff mechanism, which is explained later on. The initial state $v_0$ is ${\sf start}$.
    \end{enumerate}
    \item The set of agents is given by $[n+1]$, i.e., one more than the number of cells that the machine $M$ uses. Recall that the first $n$ agents are called ``simulators'' since their sole purpose is to simulate the computations of $M$ by tracking its IDs. The last agent, Agent $n+1$, has a more complicated role. At a high level, this agent chooses between simulating a computation of $M$ and choosing its existential transitions versus taking a fixed payoff. The idea behind this reduction is that this agent is taking the fixed payoff to start and is incentivized to simulate a computation of $M$ iff it can create an accepting computation tree.
    \item For $i \in [n]$, the the reachability goal $G_i$ is the state set $V$. For $ i = n+1$, the reachability set is $\{ { \sf goal } \}$.
    \item The playing function $P$ is specified in the following way:
    \begin{enumerate}
        \item A state in $v \in V_1$ labeled by $ \langle i, \emptyset \rangle$, for $i \in[n]$, has $P(v) = \{i\}$. This corresponds to the head being over cell $i$ in a non-existential state, and so the game is awaiting the transition information from Agent $i$.
        \item A state in $v \in V_1$ labeled by $ \langle i, \vee \rangle$ for $i \in [n]$ has $P(v) = \{n+1\}$. This corresponds to the head being over cell $i$ in an existential state, and so the game is awaiting the existential choice from Agent $n+1$. 
        \item A state $v \in V_2$ has $P(v) = \emptyset$. Such states are uncontrolled, as they just ``announce" the transition information for all agents to see and then transition to the appropriate state in $V_1$ that corresponds to the new head position and whether the new state is existential or not.
        \item A state in $v \in V_3$ labeled by $ \langle i, \alpha \rangle$ or $\langle i, \beta \rangle$ has $P(v) = \{i\}$. Recall that states in $V_3$ are visited when the game is previously in a state of type $\langle i , \vee \rangle$ and awaiting input from Agent $n+1$. Once this input about the existential choice is given, Agent $i$ is once again responsible for outputting the transition information. 
        \item $P({ \sf goal }) = P({\sf base}) = P({ \sf sink}) = \emptyset$. $P({\sf start}) = \{n+1\}$. These states correspond to the fixed payoff mechanism. The ${\sf start}$ state belongs to Agent $n+1$ since it is his choice at the start of the game to simulate a computation of $M$ or take the fixed payoff. 
    \end{enumerate}
    Since we have $|P(v)| \leq 1 $ for all $v\in V$, this is a $1$-bounded concurrent game (a.k.a. \emph{turn-based}).
    \item For $i \in [n]$, the action set $A_i$ is given the set $\{ [n] \times \Gamma \times \{\rightarrow,\leftarrow \} \times R \} \cup \{*\} \cup \{X\}$.  The set $\{ [n] \times \Gamma \times \{\rightarrow,\leftarrow \} \times R \}$ corresponds to the transition information that the agents are expected to output. The character $*$ is a special ``acceptance" character that the agents output when the machine $M$ reaches an accepting state. Likewise, $X$ is a special character corresponding to rejection. 
    
    The idea here is that whenever Agent $i$ outputs an action about the transition information, the head of $M$ is over cell $i$. This means that the set $A_i$ could be represented as $\{ \{i \} \times \Gamma \times \{\rightarrow,\leftarrow \} \times R \} \cup \{*\} \cup \{X\}$ , since the only ``previous head position" Agent $i$ reports is $i$. Nevertheless, we stick with using $[n]$ as the first component to streamline the notation -- this way, all action sets $A_i$ are equal, and they are equal to $V_2$, which reinforces the intuitive connection between these action sets and $V_2$.  
    \item The action set $A_{n+1}$ is given by $\{\alpha,\beta \}$. These actions have two interpretations. The first is at the start of the game. Taking the $\alpha$ action chooses the fixed payoff, and choosing the $\beta$ action starts the simulation of $M$. If the simulation is chosen, Agent $n+1$ uses the $\alpha$ and $\beta$ actions to stipulate which existential successor they want to see at the existential states of $M$. 
    \item The transition function $\delta$ is given by the following:
    \begin{enumerate}
        \item From a state $v= \langle i, \emptyset \rangle \in V_1$ for some $i \in [n]$ and 
        an action $\langle i, g, d, r \rangle$ that belongs to the action set  $ A_i = \{ [n] \times \Gamma \times \{\rightarrow,\leftarrow \} \times R \}$,
        the transition goes to the state $\langle i, g, d, r \rangle$ in $V_2$, mirroring Agent $i$'s choice of action. This corresponds to the transition information being faithfully reproduced in the state of the game. For the action $*$, which signifies an accepting computation, the transition from $v$ goes to the ${ \sf goal}$ state, which corresponds to acceptance. For the action $X$, which corresponds to a rejecting computation,  the transition from $v$ goes to the ${ \sf sink }$ state, which corresponds to rejection. 
        \item From a state $v=\langle i, \vee \rangle \in V_1$  for some $i \in [n]$, the action $\alpha$ from Agent $n+1$ transitions $v$ to the state $\langle i, \alpha \rangle$ in $V_3$. This is defined similarly for the $\beta$ action. This is how Agent $n+1$'s choices at existential states are recorded. 
        \item From a state $v=\langle i, \alpha \rangle $ or $v=\langle i, \beta \rangle \in V_3$ and an action $\langle i,g,d,r \rangle$ in $A_i$ belonging to the set $\{ [n] \times \Gamma \times \{\rightarrow,\leftarrow \} \times R \}$, the transition from $v$ once again goes to the state $\langle i,g,d,r \rangle$ in $V_2$. For the action $*$, the transition from $v$ goes to the ${ \sf goal}$ state. For the action $X$, the transition from $v$ goes to the ${\sf sink}$ state.
        This case is analogous to the $ v = \langle i, \emptyset \rangle \in V_1$ case discussed above; this time, however, the existential choice of Agent $n+1$ is accounted for.
        \item From a state $v=\langle i, \gamma, d, r\rangle$ in   $[n] \times \Gamma \times \{\rightarrow,\leftarrow \} \times R$ in $V_2$ the transition from $v$ is given by one of three possible cases:
        \begin{enumerate}
            \item If moving the head from position $i$ in the direction $d \in \{\rightarrow, \leftarrow\}$ results in a head position that is not in $[n]$, the transition moves to ${\sf sink}$. This represents the head moving out of bounds and the computation of the machine rejecting.
            \item Otherwise, we assume the head is moved to a correct position $i'$. 
            \begin{enumerate}
                \item If the state $r$ is existential, the transition moves to $\langle i', \vee \rangle$.
                \item Otherwise, the transition moves to $\langle i', \emptyset \rangle$
            \end{enumerate}
         \end{enumerate}
         \item From the ${\sf start}$ state, the $\alpha$ action by Agent $n+1$ moves to the  ${\sf base}$ state, corresponding to Agent $n+1$ taking the fixed payoff option. The $\beta$ action by Agent $n+1$ moves to either the state $\langle 1, \vee \rangle$ if the initial state $r_0$ of the Turing machine $M$ is non-deterministic and $\langle 1,\emptyset \rangle$, otherwise. This corresponds to simulating a computation of $M$. 
         \item The uncontrolled ${\sf base}$ state transitions back to ${\sf base}$ with probability $0.25$ and transitions to the ${\sf goal}$ state with probability $0.75$.        
    \end{enumerate}
    \item The size bound $\mathbb{L} = 3$ is a constant that does not depend on the specification of the machine $M$.  The only probabilities we use in this construction and the construction of the strategy profile in Section~\ref{profileconstruction} belong to the set $\{0, 0.25, 0.5, 0.75, 1 \}$. All of these probabilities can be represented in binary using at most $3$ digits. 
    \item The finite-time horizon $F$ is given by $(3 \cdot |\Gamma \times R|)^n) +1$,  where $|\Gamma \times R| ^n$ is an upper bound on the number of valid ID strings that the machine $M$ has. Here, we make the assumption that $R$ contains some special character that represents ``no-state" in cells where the head is not located. The number $F$ is exponential in the size of the input but has a polynomial number of digits and can be computed in polynomial time through algorithms like fast exponentiation. We multiply this bound by three as a state transition in the TM $M$ may require three state transitions in the game.

\end{enumerate}
Note that all components can be constructed in polynomial time since the playing function mandates that, at most, one agent selects an action at each state. Therefore, the transition table only needs to consider at most pairs (as opposed to tuples as large as $|\Omega|$) of states and actions, giving it a polynomial-size representation with respect to the machine $M$.

\subsubsection{Input Strategy Profile Construction}\label{profileconstruction}
\begin{figure}[ht]
  \centering
  \includegraphics[width=\textwidth]{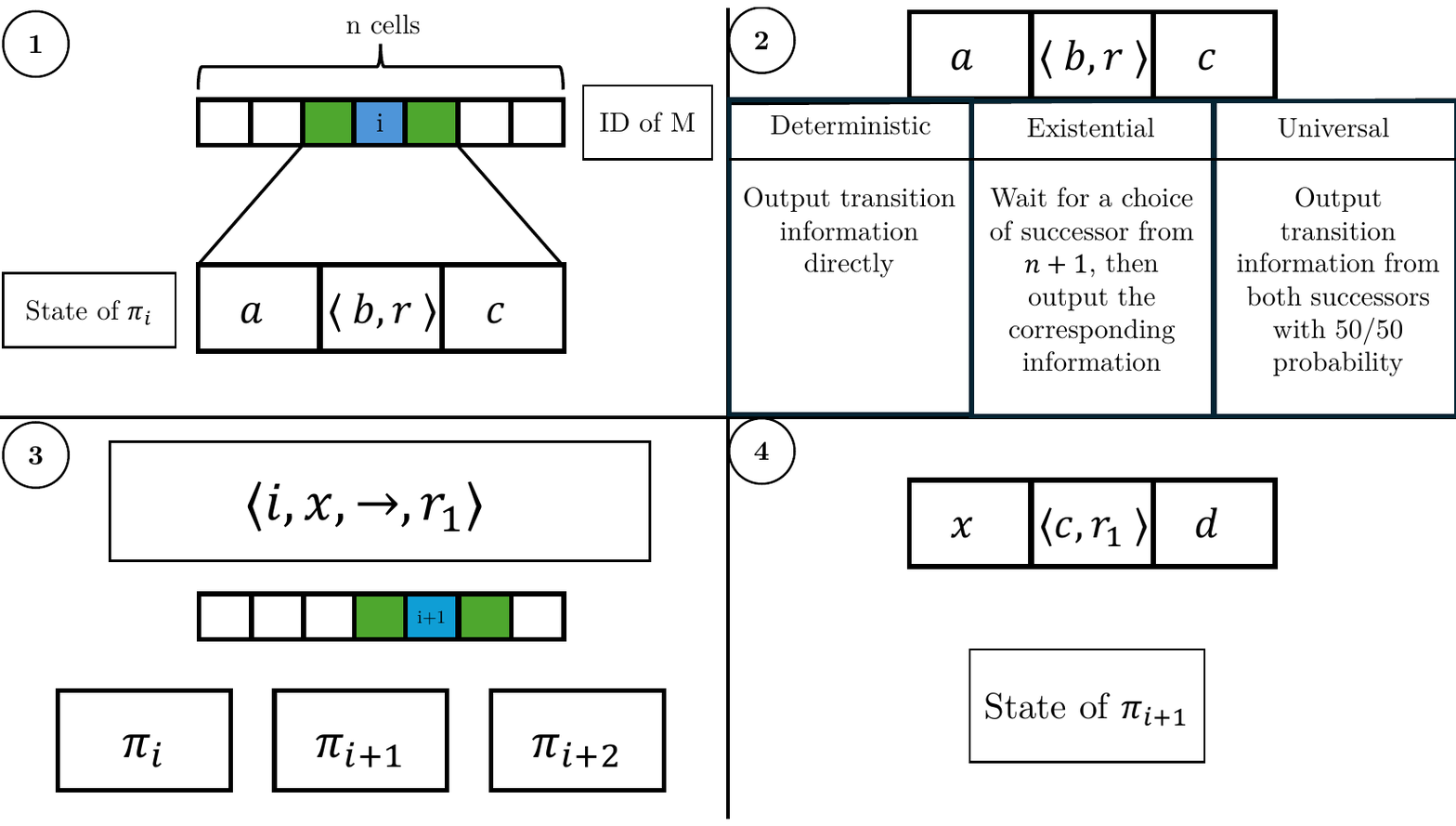}
  \caption{
  A graphical depiction of the interactions between $\pi$ and $\mathbb{G}$. As mentioned before, we use the state space of the individual $\pi_i$ to track the ID of $M$.
  {\bf (1)} The machine $M$ has some ID, depicted by the top row of small cells. The head is currently over cell $i$, which is in blue. Cells $i-1$ and $i+1$ are depicted in green. The state of $\mathbb{G}$ is $\langle i, \emptyset \rangle \in V_1$, indicating that it is Agent $i$'s turn to select an action.
  To do so, Agent $i+1$ consults the state of his assigned strategy $\pi_{i+1}$. This state records the contents of the cells $i-1,i$, and $i+1$, noting that they read $a$, $b$, and $c$ in that order. The head is currently over cell $i$ in state $r$.
  {\bf (2)} The transition information that is output by $\pi_i$ depends on the state. If $r$ is deterministic, the transition information is output in a straightforward manner. If it is existential, then Agent $n+1$ provides input indicating which successor he would like (more information on how this is managed is available in Section~\ref{profileconstruction}). If it is alternating, both successors are output with 50\% probability. 
  {\bf (3)} In this example, the transition information $\langle i, x, \rightarrow, r_1 \rangle$ is output by $\pi_i$. The game $\mathbb{G}$ then moves to the state $\langle i, x, \rightarrow, r_1 \rangle \in V_2$, which allows all agents to see how the ID has updated. In this case, the character $x$ was written at cell $i$, and the head moved right to cell $i+1$ and transitioned to state $r_1$. We highlight $\pi_{i}, \pi_{i+1}$ and $\pi_{i+2}$, as these are the strategies of the Agents that monitor cell $i+1$ and can therefore see the new position of the head at cell $i+1$. Note, however, that $\pi_{i-1}$ would also update due to the new character at cell $i$.
  {\bf (4)} We now take a look at the state of $\pi_{i+1}$ since $i+1$ is the new position of the head. This state has been updated in response to the transition information previously provided by Agent $i$. It is now Agent $i+1$'s turn to output the new transition information, effectively restarting at Step 1.
  }

  \label{fig:simulation}
\end{figure}

In this section, we provide details on how to construct the strategy profile $\pi$. Recall that the set of actions for Agents $1$ through $n$ is given by $\{ [n] \times \Gamma \times  \{\rightarrow, \leftarrow\} \times R \} \cup \{*\} \cup \{X\}$ as defined in Section~\ref{gameconstruction}. For Agent $n+1$, the set of actions is given by $\{  \alpha, \beta \}$. Before continuing with the technical details, we refer the reader to Figure~\ref{fig:simulation}, which provides a high-level overview of the interaction between $\pi$ and $\mathbb{G}$ in a graphical manner.

As mentioned before, Agent $n+1$'s assigned strategy $\pi_{n+1}$ instructs him to choose the $\alpha$ action at the ${\sf start}$ state. This gives him a probability of $1- (0.25)^{(F-1)}$ of reaching the ${\sf goal}$ state, as the only way to not reach the ${\sf goal}$ state is to loop at least $F-1$ times at the ${\sf base}$ state. The transducer implementing this strategy is of constant size.

We now consider Agents $1$ through $n$, so when we refer to $\pi_i$ in the rest of this subsection, we assume $i \in [n]$. Intuitively, the purpose of $\pi_i$ is to accurately simulate and update the ID to ID transitions that the machine $M$ makes by tracking the $i$-th character of the ID. 

Recall that a strategy $\pi_i$ is a 5-tuple $\langle S_i ,s_0^i, V, \gamma^i, O_i \rangle$.
\begin{enumerate}
    \item The state set is $S_i = \Gamma \times \{R \cup \emptyset \} \times  \{ \alpha, \beta, \emptyset \}$. As mentioned before, this state set contains complete information of the $i$-th character of the ID through the pair $\Gamma \times \{R \cup \emptyset \}$. The set $\{ \alpha, \beta, \emptyset \}$ is used to keep track of Agent $n+1$'s choices in existential successors. The $\alpha$ and $\beta$ symbols correspond to recording a choice of $\alpha$ and $\beta$ successor, respectively, while the $\emptyset$ symbol corresponds to no choice recorded.
    \item The initial state of each $\pi_i$ corresponds to the initial configuration of the tape. Since we assumed the head started in the leftmost cell, the initial state of $\pi_1$ is $\langle \emptyset, r_0, \emptyset \rangle$, corresponding to the head starting in this cell in state $r_0$. For the other $n-1$ agents, the initial state of $\pi_i$ is $\langle \emptyset,\emptyset,\emptyset \rangle$, signifying an empty cell without the head. 
    \item The input alphabet $V$ is the state set $V$ of the game $\mathbb{G}$ constructed in Section~\ref{gameconstruction}.
    \item The transition function $\gamma^i$ is based on component-wise updates.
    Let $g$ be an element of $\Gamma$, $r$ an element of $\{R \cup \emptyset \}$, and $s$ an element of  $\{ \alpha, \beta, \emptyset \}$. Assume the transducer $\pi_i$ is in state $\langle g, r, s \rangle$. We now describe how the transition function $\gamma^i$ behaves by describing how each of the three components transitions upon reading a state $v$:
    \begin{enumerate}
        \item
        If the input is $ v= \langle i, g', d, r' \rangle \in V_2$, for $i \in [n]$, $g' \in \Gamma$, $d' \in \{\rightarrow, \leftarrow\}$, and $r' \in R$.
        This corresponds to a new transition in the machine $M$ being observed, potentially updating the elements $g$ and $r$ in the state space of $\pi_i$.  The state  $v=\langle i, g', d,r' \rangle$ represents the following information about the transition in the machine $M$: the head is in cell $i$, where it writes the character $g'$ onto the tape, transitions to the new state $r'$, moving the head in direction $d$. The agents then use this information to update the state of their individual strategies $\pi_i$.  Specifically, as $\pi_i$ is in state $\langle g,r,s\rangle$:
        \begin{enumerate}
            \item 
            If $v = \langle i, g', d, r \rangle$, then this conveys the information that the character $g'$ is written in cell $i$. The first component of $\langle g,r,s \rangle$ is then updated to $g'$ to reflect the new contents of cell $i$ in the machine $M$'s ID.
            \item 
            If $v = \langle j,g',d,r \rangle$ for a cell $j$ such that either $j = i-1$ and $d = \rightarrow$ or $j = i+1$ and $d = \leftarrow $, then the head is moving to cell $i$ in state $r'$. So the middle component of $\langle g,r,s \rangle$ is then updated to $r'$ to account for the new head position and state in the machine $M$'s ID.
            
            If the new cell that results from moving from cell $j$ in direction $d$ is not $i$, then the middle component of $\langle g,r,s \rangle$ is set to $\emptyset$ to reflect the fact that the head is not positioned over cell $i$.
            \item The last component of  $\langle g,r,s \rangle$ is reset to $\emptyset$, regardless of the state $v \in V_2$ visited. Since $v \in V_2$ represents a new transition being made in $M$, the saved existential choice of Agent $n+1$ is reset.
            \end{enumerate}
        Note that multiple updates may happen simultaneously; for example, it is possible for all three components of $\langle g,r,s \rangle$ to update simultaneously when the machine writes $g'$ in cell $i$ (updating $g$), moves the head away from cell $i$ (updating $r$), and the saved value $s$ is reset.
        \item If a state in $v = \langle i, \alpha \rangle \in V_3$ is input, this corresponds to Agent $n+1$ communicating the choice of the $\alpha$ successor of an existential state in $M$. The state $\langle g,r,s \rangle$ then updates the last component to $\alpha$, putting $\pi_i$ in state $\langle g,r,\alpha \rangle$.  If the state $ v= \langle i, \beta \rangle \in V_3$ is input, the last component is analogously updated to $\beta$. If a state $v = \langle j, \alpha \rangle \in V_3$ is input for $j \not = i$, then $\pi_i$ essentially ignores this information and self-transitions, the same for $\langle j, \beta \rangle$.
        \item States in $V_1$ and the ${\sf start}$, ${\sf goal}$, ${\sf base}$, and ${\sf sink}$ states do not contain information relevant to the ID of the machine $M$, so upon an input of one of these states, the state $\langle g,r,s \rangle$ of $\pi_i$ self-transitions back to $\langle g,r,s \rangle$.
                \end{enumerate}
    Note that this transition function is deterministic.
    \item Recall that the output function had a domain of type $S_i \times V$ solely so that the element $v \in V$ could suppress the output of $\pi_i$ when $ i \not \in P(v)$. We, therefore, describe the output function based solely on the state $\langle g, r ,s \rangle$, where we once again have $g \in \Gamma, r \in R, s \in \{\alpha,\beta, \emptyset\}$.
    \begin{enumerate}
     \item  If $\kappa(r) = \mathrm{accept}$, then $r$ is an accepting state, and  the output is *.         If $\kappa(r) = \mathrm{reject}$, then $r$ is a rejecting state, and the output is $X$.
    \item If $\kappa(r) = -$, then $r$ is a deterministic state. The output at $\langle g,r,s \rangle$ is then the action corresponding to the transition of $M$. For example, if the transition is to write the character $g'$, move the head to the left, and transition to state $r'$, then the output $\langle i, g', \leftarrow, r' \rangle$. 
    \item If $\kappa(r) =\wedge$,  then $r$ is a universal state and the output at state  $\langle g,r, s\rangle$ is a $50/50$ distribution over the actions that correspond to the transitions for  $r_{\alpha}$ and $r_{\beta}$, where the actions that correspond to $r_{\alpha}$ and $r_{\beta}$ are analogous to the deterministic setting above.
        Note that for both universal and deterministic states, we define the output function at all values of $s$ for completeness. In the actual run of the transducers, however, the only element of $s$ we see at these states is $\emptyset$, as neither deterministic nor universal states register an existential successor choice from Agent $n+1$. 
        \item If $\kappa(r) = \vee$, then $r$ is an existential state. Note that Agent $i$ does not actually play immediately at existential states, as they must wait for input from Agent $n+1$. So at states $\langle g,r, \emptyset \rangle$, the output is inconsequential as the transition function $\delta$ of $\mathbb{G}$ ultimately ignores it. At the state $\langle g,r, \alpha \rangle$, the output corresponds to the $\alpha$ successor transition of $r$,
        $r_{\alpha}$, and analogously for $\beta$. 
        \end{enumerate}
    \end{enumerate}
An easy argument by induction shows that if the game is in a state $v$ such that $i \in P(v)$, then the head is in position $i$ in $M$. This can be seen by simply noting that the $\mathbb{G}$ accurately tracks the head through its transition function $\delta$ and state space $V$. Agent $i$ is then only asked to output a relevant action whenever the head is over cell $i$. Therefore, the output at states of the form  $\langle g, \emptyset,s \rangle$ is irrelevant and can be thought of as an arbitrary action choice.

Taken together, since each $S_i$ accurately reflects the value of the $i$-th character of $M$'s ID and transitions correctly in $\pi_i$ upon receiving information about the current transition in $M$, the state spaces of the transducers $\pi_i$ accurately reflect the IDs of the machine $M$.

Note that all components of $\pi_i$ can be constructed in polynomial time w.r.t the input. The state space is a cross product of $\Gamma$ and $R$ and an additional set of constant size. The output function associates these states to distributions over a polynomially-sized action set that contains at most two elements. The transition function consists of triples $\langle s_1, v, s_2 \rangle$ for $s_1,s_2 \in S_i$ and $v \in V$. Both $S_i$ and $V$  are polynomial in size w.r.t the input, so the set of all triples   $\langle s_1, v, s_2 \rangle$ (an accurate representation of the deterministic transition function $\gamma^i)$ is also polynomial in the size of the input. 

With both $\mathbb{G}$ and $\pi$ constructed, we move on to proving the validity of the reduction.

\subsubsection{Validity of the Reduction}

Before proving the validity of the reduction directly, we first require an observation regarding the size of accepting computation trees of $M$. Let the space-bounded machine $M$ have at most $B$ possible valid IDs and accept the empty tape, meaning that it has a valid computation tree $T$. Then, it has an accepting computation tree $T'$ in which all of its branches are limited in length by $B$. In order to see this, consider a valid computation tree $T'$ with a branch that has more than $B$ elements. This branch must be finite since it represents an accepting computation that must halt. By the pigeonhole principle, at least one ID $I$ must be repeated on this branch. By replacing the subtree rooted at the first occurrence of this ID on the branch with the subtree rooted at the last occurrence on the branch, we create a new accepting computation tree $\hat{T}$ where $I$ is no longer repeated on the branch. By repeating this process, we can create a computation tree where no ID is repeated within a branch.   

\begin{rem}\label{sizeoftree}
    Let $M$ be a space-bounded alternating Turing Machine with at most $B$ possible valid IDs. Assume that $M$ accepts the empty tape, meaning that it admits a valid computation tree $T$. Then, it admits a valid computation tree $T'$ in which each branch of the tree has a length at most $B$.
\end{rem}

We now state our main theorem.
\begin{thm}\label{valid}
    The strategy profile $\pi = \langle \pi_1 \ldots \pi_{n+1} \rangle$ is an NE in $\mathbb{G}$ iff $M$ does not accept the empty tape using at most $n$ cells.
\end{thm}

\begin{proof}
    
Recall that the input strategy $\pi_{n+1}$ uses the $\alpha$ action at all time steps. Therefore, the input strategy profile gives Agent $n+1$ a payoff of $1- (0.25)^{(F-1)}$, as it chooses $\alpha$ at the start state to transition to the base state.  The simulating agents $ i \in [n]$ receive a payoff of $1$ no matter what and, therefore, have no incentive to deviate. Therefore, checking that $\pi$ is a Nash equilibrium amounts to checking if Agent $n+1$ has a profitable deviation, i.e., a strategy that gives him a payoff larger than $1- (0.25)^{(F-1)}$.

Suppose that $M$ accepts the empty tape. Then, it admits an accepting computation tree $T$. By Remark~\ref{sizeoftree}, it also accepts a computation tree $T'$ in which no branch contains a repeated ID. By following the strategy outlined by $T'$ (choosing the same existential transition at an ID as $T'$), Agent $n+1$ can meet his goal with probability $1$ in less than $F-1$ steps. Since $1 > 1- (0.25)^{(F-1)}$, Agent $n+1$ has an incentive to deviate from $\pi_{n+1}$ and $\pi$ is not a Nash equilibrium.

Now, suppose that $M$ does not accept the empty tape, meaning that it has no accepting computation tree. Therefore, regardless of the choice of the successors at the existential states, every computation tree has at least one rejecting branch. The simulations of $M$ in $\mathbb{G}$ probabilistically branch at the universal states, meaning that the simulation of $M$ in $\mathbb{G}$ probabilistically samples branches of the computation tree of $M$.   A lower bound on the probability that a non-accepting branch is sampled is given by $(0.5)^{F-1}$ under the overly strong assumption that every transition in $M$ corresponds to a universal state that must pick a single specific sequence of choices in order to reach a rejecting branch (this assumption is overly strong as it assumes that each transition in $\mathbb{G}$ corresponds to a new transition in $M$, which is not the case). Note that if there are fewer universal choices, the probability becomes larger; if there are no universal choices, this probability is $1$. This gives us an upper bound of $1-(0.5)^{F-1}$ on Agent $n+1$'s payoff for a non-accepting $M$, as a non-accepting branch is sampled with at least probability $(0.5)^{F-1}$ in the simulation of $M$ by $\mathbb{G}$ under the assumption that $M$ does not have an accepting computation tree. This means that Agent $n+1$ does not have an incentive to deviate as we have $1-(0.5)^{F-1} < 1-(0.25)^{F-1}$, so $\pi$ is a Nash equilibrium.
\end{proof}

Theorem~\ref{valid} gives us our desired lower bound.

\begin{thm}\label{lower}
The problem of verifying whether a given strategy profile $\pi$, represented by a tuple of
probabilistic transducers, is a Nash equilibrium in a given $b$-bounded probabilistic concurrent game  $\mathbb{G}$ is EXPTIME-complete. 
\end{thm}

\section{Concluding Remarks and Future Work}

The two focal results of this paper are Theorem~\ref{upper}, which establishes a PSPACE upper bound for the subgame-perfect-equilibrium verification problem, and Theorem~\ref{lower}, which establishes an EXPTIME lower bound for the Nash-equilibrium verification problem. These results contradict our intuitive understanding of the two solution concepts, as the subgame-perfect equilibrium is seen as a stricter refinement of the Nash equilibrium that offers conceptual benefits at the cost of being harder to reason about.   Here, we see that for the verification problem specifically, the stricter conditions of the subgame-perfect equilibrium make it easier to reason about than the Nash equilibrium by allowing for efficient parallelization. When we contrast this with the fact that Nash equilibria are usually computed in these systems through the same backwards induction algorithm that creates subgame-perfect equilibria, these two observations paint an interesting picture of the two equilibrium concepts that we intend to investigate in future work. For example, if we consider infinite-horizon systems, does this relationship still hold? If so, this strengthens the evidence for the conjecture that the subgame-perfect equilibrium is a strictly preferable solution concept in probabilistic systems. 

If we consider a deviating agent $j$, we can see the fundamental difference between subgame-perfect equilibrium and Nash equilibrium -- by examining the Markov Decision Process $\mathbb{G}_i \times \pi$ (see Section~\ref{improve}), we can see that the conditions of a strategic improvement for agent $j$ are different. For the subgame-perfect equilibrium, if the value at some history-relevant state can be improved, then $\pi_j$ can be deviated from, and $\pi$ is not a subgame-perfect equilibrium. This closely corresponds to the concept of an ``optimal policy" from the literature on Markov Decision Processes~\cite{Puterman94}, which is a memoryless strategy that maximizes the value of all states simultaneously. For the Nash equilibrium, however, the agent is only incentivized to deviate if the value of the initial state specifically can be improved, which does not directly correspond to notions of optimality in the Markov-Decision-Process literature. This observation paints an interesting picture of the relationship between these equilibrium concepts and other strategy-related solution concepts. 

In this paper, we considered a model of concurrent games with an explicitly specified bound on the number of agents that can act concurrently at any given time. As mentioned before, this was done to improve the accuracy of the complexity-theoretic analysis. If we allowed for a fully concurrent system, we would still get the same PSPACE upper bound result of Theorem~\ref{upper} and EXPTIME upper bound result of Theorem~\ref{nashupper}, but they would both lose some of their significance due to the fact that $\delta$ itself can be seen as an exponential construction on its own.  For example, for the subgame-perfect equilibrium, the question would then be if the final complexity result was PSPACE with respect to the encoding size of the large transition table and, therefore, practically doubly-exponentially in the number of agents.  This would have yielded a weaker result than the one we show here, so we took special care to introduce and consider $b$-bounded systems, which we knew would admit a smaller representation -- a very common theme in the field of multi-agent systems.  While we can establish the same upper bounds for fully concurrent systems as for $b$-bounded ones in this paper, properly considering complexity-theoretic results for fully concurrent systems requires \emph{multivariate analysis} to be meaningful. A multivariate analysis would be necessary to characterize the complexity in terms of the size of the transition table independently of the number of agents in order to better understand what is actually influencing the complexity-theoretic results presented. We leave this multivariate analysis for future work.

\FloatBarrier
\bibliographystyle{alphaurl}
 \bibliography{bib}
\end{document}